\documentclass[webpdf,modern,medium, namedate]{oup-authoring-template}

\onecolumn 

\graphicspath{{Fig/}}

\theoremstyle{thmstyleone}%
\theoremstyle{thmstyletwo}%
\theoremstyle{thmstylethree}%

\usepackage{rotating}  
\usepackage{pdflscape}   
\usepackage{booktabs}    
\usepackage{siunitx}    
\begin{document}

\journaltitle{Journal Title Here}
\DOI{DOI HERE}
\copyrightyear{2022}
\pubyear{2019}
\access{Advance Access Publication Date: Day Month Year}
\appnotes{Paper}

\firstpage{1}


\title[A decay-adjusted spatio-temporal model for prevalence mapping]{A decay-adjusted spatio-temporal model to account for the impact of mass drug administration on neglected tropical disease prevalence}

\author[1,2,$\ast$]{Emanuele Giorgi}
\author[1]{Claudio Fronterre}
\author[2]{Peter J Diggle}

\authormark{Giorgi, E., Fronterre, C. and Diggle, P. J.}

\address[1]{\orgdiv{BESTEAM, Department of Applied Health Sciences}, \orgname{University of Birmingham}, \postcode{Birmingham}, \country{United Kingdom}}
\address[2]{\orgdiv{CHICAS, Lancaster Medical School}, \orgname{Lancaster University},  \postcode{Lancaster},\country{United Kingdom}}

\corresp[$\ast$]{Corresponding author. \href{email:e.giorgi@bham.ac.uk}{e.giorgi@bham.ac.uk}}

\received{Date}{0}{Year}
\revised{Date}{0}{Year}
\accepted{Date}{0}{Year}



\abstract{Prevalence surveys are routinely used to monitor the effectiveness of mass drug administration (MDA) programmes for controlling Neglected Tropical Diseases (NTDs). We propose a decay-adjusted spatio-temporal (DAST) model that explicitly accounts for the time-varying impact of MDA on NTD prevalence, providing a flexible and interpretable framework for estimating intervention effects from sparse survey data. Using case studies on soil-transmitted helminths and lymphatic filariasis, we show that DAST offers a practical alternative to standard geostatistical models when the objective includes quantifying MDA impact and supporting short-term programmatic forecasting. We also discuss extensions and identifiability challenges, advocating for data-driven parsimony over complexity in settings where the available data are too sparse to support the estimation of highly parameterised models.}



\maketitle

\section{Introduction}\label{sec:introduction}
The term \textit{Neglected Tropical Diseases} (NTDs) was introduced in 2005 to describe a diverse group of infectious diseases found primarily in tropical and subtropical regions. These diseases are most prevalent in resource-limited communities and have historically attracted insufficient investment in research, treatment, and vaccine development, as well as limited political prioritisation of their control and elimination. Although NTDs rarely cause death, they often lead to long-term disabilities, such as blindness, disfigurement, and chronic fatigue, which severely impact individual quality of life and economic productivity. The control of NTDs is guided by three fundamental strategies: prevention, accurate diagnosis, and effective treatment.

One of the most widely implemented strategies for controlling NTDs is mass drug administration (MDA), which involves the regular distribution of medications to entire at-risk populations, regardless of individual infection status. As a preventive public health intervention, MDA aims to reduce both individual infection and community-level transmission, and has become a cornerstone of global efforts to control and eliminate several major NTDs, such as schistosomiasis, soil-transmitted helminthiases (STH), lymphatic filariasis, onchocerciasis, and trachoma. Although caused by diverse pathogens, ranging from parasitic worms to bacteria, this group of NTDs all respond effectively to MDA, whose impact can be observed within a range of days to months since its roll out. However, sustained community-level benefits and potential interruption of transmission typically require repeated MDA cycles over years (e.g. \cite{anderson2015mda}). Furthermore, the drugs used in MDA typically cause only mild and transient side effects, such as gastrointestinal discomfort or dizziness, or no side effects at all, making them safe for administration on a large scale, even to asymptomatic individuals \citep{romani2018}.

Model-based geostatistics (MBG) \citep{diggle1998} has become an established likelihood-based framework for disease prevalence mapping, particularly in low-resource settings. In MBG, disease prevalence is modelled as a function of explanatory variables (often derived from remotely sensed environmental and socio-demographic data) and a residual spatial Gaussian process that accounts for unmeasured spatial heterogeneity. This framework has been widely applied in the context of neglected tropical diseases (NTDs) to inform resource allocation and guide targeted interventions. Notable examples include mapping efforts for schistosomiasis, lymphatic filariasis, and trachoma in sub-Saharan Africa and Southeast Asia (e.g. \citet{hay2013globalmapping, pullan2014global, diggle2016mbg, sasanami2023}).

While MBG has been applied to combine data collected at different stages of intervention, most current approaches either ignore the effect of MDA or include it only as a covariate, typically represented by the cumulative number of rounds administered \citep{schmidt2022, aye2018}. This simplification does not account for the timing of MDA, which is critical to capturing its dynamic impact on infection prevalence. The effect of MDA unfolds gradually, with reductions in prevalence occurring over time following each round. In some cases, particularly when survey data are sparsely distributed in space or when follow-up surveys are more likely to target areas with high residual transmission, using MDA as a standard covariate in a regression model can lead to biased estimates of its association with prevalence. In turn, this can result in MDA appearing to be ineffective, due not to lack of impact, but to limitations in how its effect is represented in a model.

The focus of this paper is to develop a novel spatio-temporal geostatistical model that explicitly incorporates the impact of MDA, while accounting for its timing, intensity, and cumulative effect. This approach aims to overcome the limitations of standard methods that treat MDA simplistically as a covariate with a linear effect, which can result in misleading or counterintuitive associations due to spatial and temporal confounding. We argue that the proposed approach is especially beneficial when combining data collected at various stages of the monitoring and evaluation process, from baseline (i.e. prior to the implementation of MDA) to follow-up surveys designed to assess changes in prevalence over time. Specifically, our goal is to define a functional form that captures the lagged impact of MDA in an interpretable and biologically plausible way, recognizing that reductions in prevalence may not occur immediately and that the influence of past MDA rounds diminishes over time but may also accumulate. This allows us to model the dynamics of disease response to intervention more realistically.

The proposed approach introduces additional structural assumptions regarding how MDA influences prevalence while retaining a strong level of predictive performance compared with standard approaches. Moreover, it enhances the interpretability of geostatistical models by constraining the effect of MDA within epidemiologically plausible temporal dynamics, thereby improving both scientific insight and practical relevance for decision-making. Hence, we argue that the proposed model has a broader applicability than standard approaches, as it can be used not only for retrospective analysis but also for forecasting prevalence under different control scenarios, making it a viable data-driven alternative to mathematical mechanistic models commonly used for this purpose (see, for example, \citet{griffin2010reducing}, \citet{stolk2013modeling}, \citet{truscott2016soil} and \citet{Touloupou2024-gf}).

In Section~\ref{sec:issues}, we review existing approaches for incorporating MDA history into geostatistical models and highlight the methodological limitations inherent to these strategies. Section~\ref{sec:methods} introduces the proposed spatio-temporal geostatistical model, detailing how the impact of MDA on disease prevalence is represented through a parameterised approximation to the underlying intervention effect. In Section~\ref{sec:simulation}, we present a simulation study designed to evaluate the predictive accuracy and robustness of the model at varying levels of spatial and temporal data sparsity. Section~\ref{sec:applications} demonstrates the application of the method to two case studies on NTDs. Finally, Section~\ref{sec:discussion} summarises the strengths and limitations of the proposed modelling framework and outlines potential extensions for more complex data scenarios beyond those considered in this paper.

\section{Issues in combining survey data across time for neglected tropical diseases}
\label{sec:issues}

Routine monitoring for NTDs is typically based on cross‐sectional prevalence surveys conducted at key programmatic time points: an initial baseline survey, undertaken before the first round of MDA, followed by one or more impact surveys to assess whether further rounds of MDA are necessary or whether elimination thresholds have been reached. These surveys are generally implemented at the community or school level, within pre-defined \emph{Implementation Units} (IUs), which serve as the spatial units for decision-making.

Due to limited logistical resources and, in some regions, security constraints, surveys are rarely conducted simultaneously across all IUs. Instead, different IUs are visited at different time points, so that consecutive samples are not available for most areas. This creates a misalignment between the spatial and temporal dimensions of the data, which complicates the interpretation of prevalence trends over time. Standard analytical approaches often assume that observed changes over time reflect underlying epidemiological trends. However, when sampling locations differ across survey rounds this assumption becomes questionable.

Such mismatches can induce systematic spatial confounding, particularly when MDA roll-out and follow-up surveys are preferentially targeted towards higher-burden areas. This confounding can lead to biased estimates of the impact of the intervention. In extreme cases, this could even lead to the misleading conclusion that prevalence is rising over time in response to MDA, when in reality it is decreasing across all IUs.

\begin{figure}[t]
  \centering
  \includegraphics[width=\textwidth]{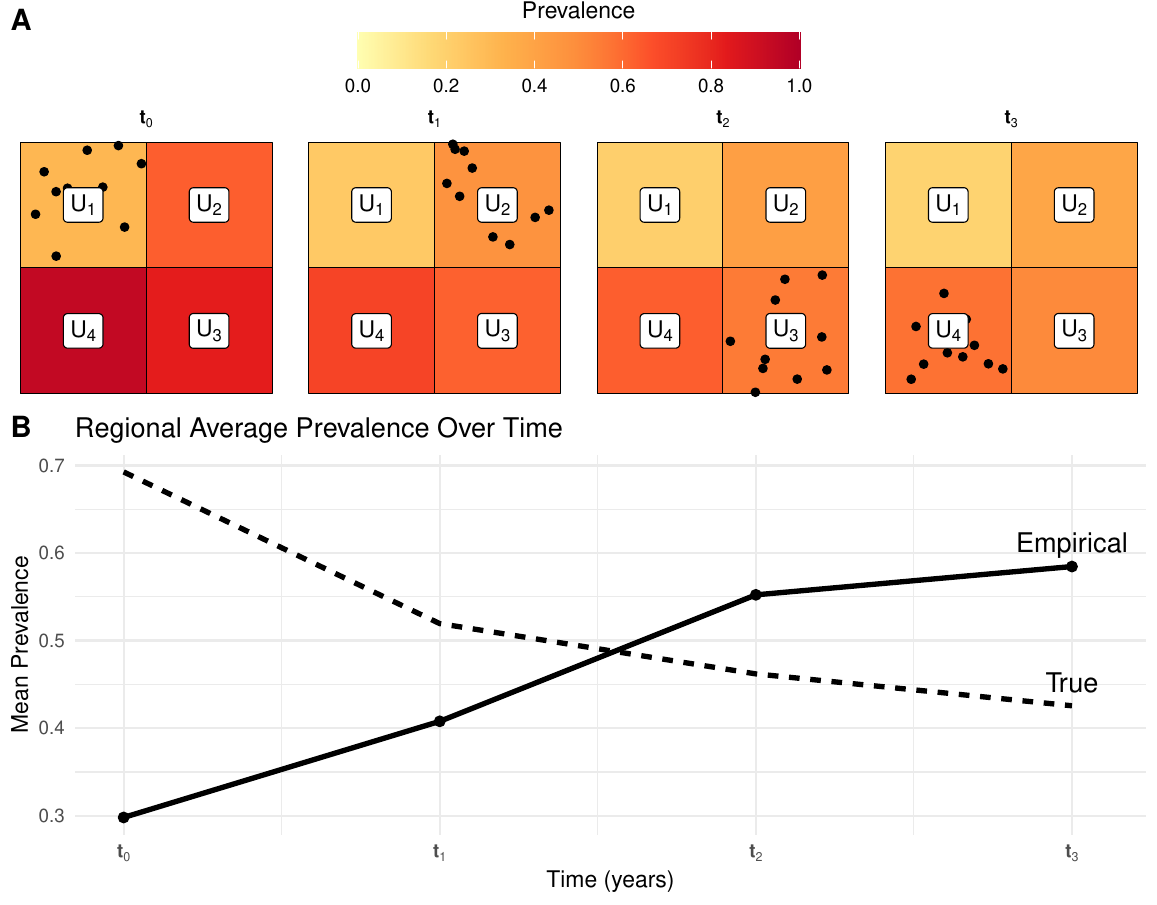}
  \caption{Schematic representation of a spatially confounded survey design and its effects. Panel A: spatial layout of IUs and sampled locations over four time points. Panel B: comparison of empirical mean prevalence at sampled locations versus true regional average over time. Although prevalence decreases in all IUs, the empirical average increases due to preferential sampling of higher-burden areas at later time points.}
  \label{fig-overview}
\end{figure}

Figure \ref{fig-overview} illustrates this issue using a stylised example. The study region is a unit square divided into four equally sized implementation units ($U_1$ to $U_4$), each with a different baseline prevalence. Within each IU, prevalence declines smoothly over time due to the cumulative effect of MDA. However, only one IU is surveyed at each time point, and the IU selected at time $t+1$ always has a higher baseline prevalence than the one sampled at time $t$. Panel $A$ shows the spatial distribution and the temporal decline within each IU. Panel $B$ displays the average prevalence in the sampled locations (solid line) alongside the true regional average in all IUs (dashed line). Despite every IU experiencing a true decline, the empirical average increases misleadingly over time. This illustrates how spatio-temporal confounding can arise from the interaction between spatial heterogeneity in baseline prevalence and temporal reductions due to treatment, particularly when survey locations shift over time.

This example is deliberately extreme, but reflects a pattern that is not uncommon in NTD prevalence data. In the absence of appropriate statistical adjustment, such survey designs can produce highly misleading conclusions about both intervention effectiveness and underlying disease dynamics. In Section \ref{sec:current_geo} we review current approaches for incorporating MDA effects in geostatistical and regression-based models. In 
Section \ref{subsec:survey_conf} we return to the example in Figure~\ref{fig-overview} to illustrate a scenario in which treating MDA history as a standard covariate leads to identifiability issues.

\subsection{Current approaches in geostatistics and other regression-based models to account for the impact of MDA}
\label{sec:current_geo}
Let $Y_i$ denote the number of infected individuals among $m_i$ examined
at location $x_i \in \mathbb{R}^2$ and time $t_i$.  We assume
\begin{equation}
  Y_i \mid P(x_i,t_i) \;\sim\;
  \text{Bin}\!\bigl(m_i,\;P(x_i,t_i)\bigr),
  \qquad i = 1,\dots,n,
\end{equation}
where $P(x,t)=\Pr\{\text{positive test for an individual at }(x,t)\}$.
The canonical formulation for geostatistical modelling of disease prevalence is
\begin{equation}
  \log\left\{\frac{P(x_i,t_i)}{1-P(x_i,t_i)}\right\}
  \;=\;
  \mathbf d(x_i,t_i)^{\!\top}\boldsymbol\beta
  + S(x_i,t_i),
  \label{eq:linpred}
\end{equation}
where $\mathbf d(x_i,t_i)$ is a vector of spatio-temporal covariates, $\boldsymbol\beta$ is a vector of fixed effects, and $S(x_i,t_i)$ is a zero-mean Gaussian process capturing residual spatial and temporal correlation due to unmeasured or unobserved factors. If the data allow, an unstructured random effect $Z_i$ can be added to the linear predictor to capture spatial variation, at scales smaller
than the minimum observed distance between sampling locations. The interpretation of $Z_i$ depends on the context of the analysis; it may reflect residual local heterogeneity not accounted for by $S(x_i,t_i)$, or account for extra-binomial variation arising from genetic, behavioural, or lifestyle-related factors.

The covariates $\mathbf d(x_i, t_i)$ typically include environmental and climatic variables such as land surface temperature, vegetation indices (e.g.\ NDVI or EVI), precipitation, and elevation. For vector-borne diseases, these covariates are commonly used as proxies for the spatial distribution and seasonal activity of the disease vector, under the assumption that transmission risk is modulated by ecological conditions favourable to vector survival and reproduction. Socio-demographic variables such as population density, access to sanitation, or socioeconomic status can also play a key role by influencing human–vector contact patterns, exposure risk, and susceptibility to infection. However, socio-economic factors are not easily captured through remotely sensed data, and are therefore often approximated using proxies such as population density or night-time light emission, which have been shown to correlate with human activity and infrastructure development (e.g. \citet{Linard2012PopulationDistribution, Mellander2015NightTimeLight}). However, the temporal and spatial sparsity of survey data can make it difficult to estimate the temporal correlation structure  in $S(x,t)$ reliably. In such cases, a practical alternative is to assume that the residual variation can be adequately captured by a purely spatial process $S(x)$. Importantly, for many neglected tropical diseases, which tend to have chronic manifestations and slow natural dynamics, temporal variation in prevalence is minimal in the absence of MDA or other major interventions. Therefore, in contexts where MDA is not ongoing, a model with a purely spatial random effect $S(x)$ may offer a reasonable approximation of the underlying disease process. We will revisit this point later when evaluating model choices for the case studies.

A pragmatic approach to defining the spatio-temporal covariance 
function for $S(x, t)$ in disease mapping applications is to assume a separable 
structure,
$$
\operatorname{Cov}\!\bigl\{S(x,t),S(x',t')\bigr\}
  = \sigma^2\,
    \rho_S\!\bigl(\lVert x-x'\rVert;\,\phi_S\bigr)\;
    \rho_T\!\bigl(\lvert t-t'\rvert;\,\phi_T\bigr),
$$
where $\sigma^2$ is the marginal variance, and $\rho_S$ and $\rho_T$ are 
exponential correlation functions with range parameters $\phi_S$ and $\phi_T$, 
respectively:
\begin{equation}
    \label{eq:st_cor}
    \rho_S(h;\phi_S) = \exp(-h/\phi_S),
\qquad
\rho_T(u;\phi_T) = \exp(-u/\phi_T),
\end{equation}
where $h = \lVert x - x' \rVert$ and $u = |t - t'|$. This double exponential 
specification is computationally efficient and avoids the identifiability 
difficulties that arise with more flexible alternatives, such as the Mat\'{e}rn 
class or the non-separable structures of \citet{gneiting2002, gneiting2007}, 
whose additional parameters (including spatial and temporal smoothness, and space--time 
interaction terms) can be poorly identifiable from the sparse and 
irregularly distributed count data 
\citep{giorgi2018}.

Based on this spatio-temporal model, a commonly used approach to account for the effect of MDA on disease prevalence is to use the cumulative number of effective MDA rounds administered prior to time $t$ at location $x$, or some transformation of this quantity (e.g. \citet{aye2018}). Formally, this covariate can be defined as
\begin{equation}
\label{eq:cum_mda}
c(x,t) = \sum_{j \: : \: u_j < t} \mathbb{I}(x, u_j) 
\end{equation}
the $ u_j $ are the times at which effective rounds of MDA occurred before time $ t $, and $ \mathbb{I}(x, u_j) = 1 $ if location $ x $ falls within the IU receiving the $ j $-th effective round of MDA, and 0 otherwise. An MDA round is typically deemed effective if it reaches a minimum threshold of population coverage, which varies by disease in accordance with WHO guidelines. For example, in the case of lymphatic filariasis, an MDA round is considered effective if at least 65\% of the total population in the implementation unit is treated with the recommended drug regimen \citep{WHO2011LF}. Similar thresholds exist for other diseases: for instance, soil-transmitted helminth control programmes commonly use 75\% coverage of the at-risk population, typically preschool- and school-aged children, as the benchmark for an effective round \citep{World-Health-Organization2024-kn}. Although the covariate $c(x,t)$ is here specified as a spatially continuous function, in practice it is almost always available only as a piecewise-constant covariate that takes a single value for all locations $x$ falling within the same IU. This discretisation reflects how MDA coverage is recorded and reported, typically aggregated at the IU level rather than measured at fine spatial resolution. This introduces an additional challenge for inference on the effect of $c(x,t)$, as its spatial resolution does not match that of the outcome variable in geostatistical modelling. As we will see, this limitation also carries forward into the approach introduced in this paper.

Several studies have explored the inclusion of MDA in regression models for NTD-related health outcomes. These vary in complexity and context, but generally treat MDA as an external covariate without  constraints, as in the following examples.

\citet{crellen2016} modelled egg reduction rates (ERRs) using a hierarchical Bayesian model with a treatment indicator interacting with covariates including MDA exposure. The cumulative number of prior MDA rounds at each school was treated as a categorical covariate influencing the magnitude of ERR. 

\citet{aye2018} developed a generalized linear model for log-transformed microfilarial prevalence, considering metrics from the MDA program, including the total rounds delivered, the time since the most recent round, and measures of program fragmentation such as the average interval between rounds; however, only baseline prevalence and an interaction between the most recent microfilariae value and the number of rounds since that measurement were retained in the final model, as the fragmentation covariates did not significantly improve model fit.

\citet{loukouri2020} employed mixed-effects models to assess differences in hookworm prevalence under annual versus semi-annual MDA. Treatment frequency entered the model as a categorical variable (annual vs semi-annual), with village-level unstructured random effects accounting for clustering.

\cite{srivathsan2024} used the number of previous MDA rounds as one of several covariates in fixed-effects and mixed-effects regression models to forecast trachoma prevalence. This variable was used alongside time and baseline prevalence in ensemble-based predictive models. 

In summary, while these studies reflect an increasing effort to incorporate programme history into prevalence modelling, the role of MDA is generally captured through simple covariates such as the number of previous rounds or cumulative years of treatment. Although such models may include more nuanced information (e.g.\ frequency, timing, or intensity of MDA), the effect of MDA is left unconstrained and estimated purely from the data. These studies did not report counterintuitive estimates of MDA effects, which may be attributed to the good spatial and temporal coverage of the areas under investigation. In contrast, other studies focusing primarily on disease risk mapping often omit MDA effects altogether, which may reflect reporting bias, as models yielding implausible or unexpected estimates are likely excluded from the presented results rather than explicitly discussed.

The central problem addressed in this paper is how to incorporate MDA history into a geostatistical model when survey data are collected using sampling strategies that do not revisit the same set of locations over time. This challenge is compounded when MDA efforts are preferentially targeted to locations with higher baseline prevalence. In such settings, treating the effect of a covariate such as $c(x,t)$ (see equation~\eqref{eq:cum_mda}) as an unconstrained term in the linear predictor can lead to biased estimates and, more problematically, to epidemiologically implausible or counterintuitive results.

To illustrate the severity of this issue, the next section presents a synthetic example in which $c(x,t)$ is perfectly confounded with spatial variation in baseline prevalence, rendering its effect non-identifiable. We then propose a 
modelling approach that constrains the impact of MDA to follow a plausible intervention dynamic. 

\subsection{Inference under confounding by indication: a simplified example}
\label{subsec:survey_conf}
To illustrate the limitations of using cumulative rounds of effective MDA as an unconstrained covariate, we consider a simplified data scenario aligned with the stylized design shown in Figure~\ref{fig-overview}. The study region is partitioned into four IUs, denoted by $U_k$ for $k = 1, \ldots, 4$. We assume that each IU has a constant baseline prevalence given by $p_{0,k} = k/5$, such that $p(x) = p_{0,k}$ for all $x \in U_k$.

A single effective round of MDA is delivered at time $t = 0$, but only in units $U_3$ and $U_4$, which have the highest prevalence at baseline. The survey design involves collecting data from 25 locations per IU, with sampling staggered over time: data are collected in $U_1$ at $t = 0$, in $U_2$ at $t = 1$, in $U_3$ at $t = 2$, and in $U_4$ at $t = 3$.

We define the time-varying prevalence in IU 3 and 4, at time $t$, following an MDA round at time $t=0$ as
\begin{equation}
\label{eq:model_sim_example}
    p_{k,t} = p_{0,k} \exp(-t/\gamma), k=3, 4,
\end{equation}
where  $\gamma$ is the decay parameter regulating the waning of MDA impact.

We now consider a binomial regression model that includes both IU-specific intercepts and the cumulative number of MDA rounds as covariates. Let $p(x, t)$ denote the probability of infection at location $x$ and time $t$, and let $\mathbb{I}_k(x)$ be an indicator function taking value 1 if $x \in U_k$ and 0 otherwise. Let $c(x, t)$ denote the cumulative number of effective MDA rounds received at location $x$ up to time $t$, previously defined in \eqref{eq:cum_mda}. The model is given by
\begin{equation}
\log\left\{ \frac{p(x, t)}{1 - p(x, t)} \right\}= \sum_{k=1}^4 \beta_k \, \mathbb{I}_k(x) + \beta_5 c(x, t),
\label{eq:glm_naive}
\end{equation}
where each $\beta_k$ captures the baseline prevalence in IU $k$, and $\beta_5$ is intended to represent the effect of cumulative MDA. However, under the survey design described above, each IU is sampled at a different time point and only once, so the cumulative MDA exposure $c(x, t)$ is perfectly collinear with the IU indicators. In particular, $c(x, t) = 0$ in $U_1$ and $U_2$, and $c(x, t) = 1$ in $U_3$ and $U_4$. As a result, $c(x, t)$ can be expressed as a linear combination of $\mathbb{I}_3(x)$ and $\mathbb{I}_4(x)$, rendering the design matrix rank deficient. Consequently, the coefficient $\beta_5$ is not estimable, and model~\eqref{eq:glm_naive} is not identifiable without further constraints.

\begin{figure}[ht]
    \centering
    \includegraphics[width=1\linewidth]{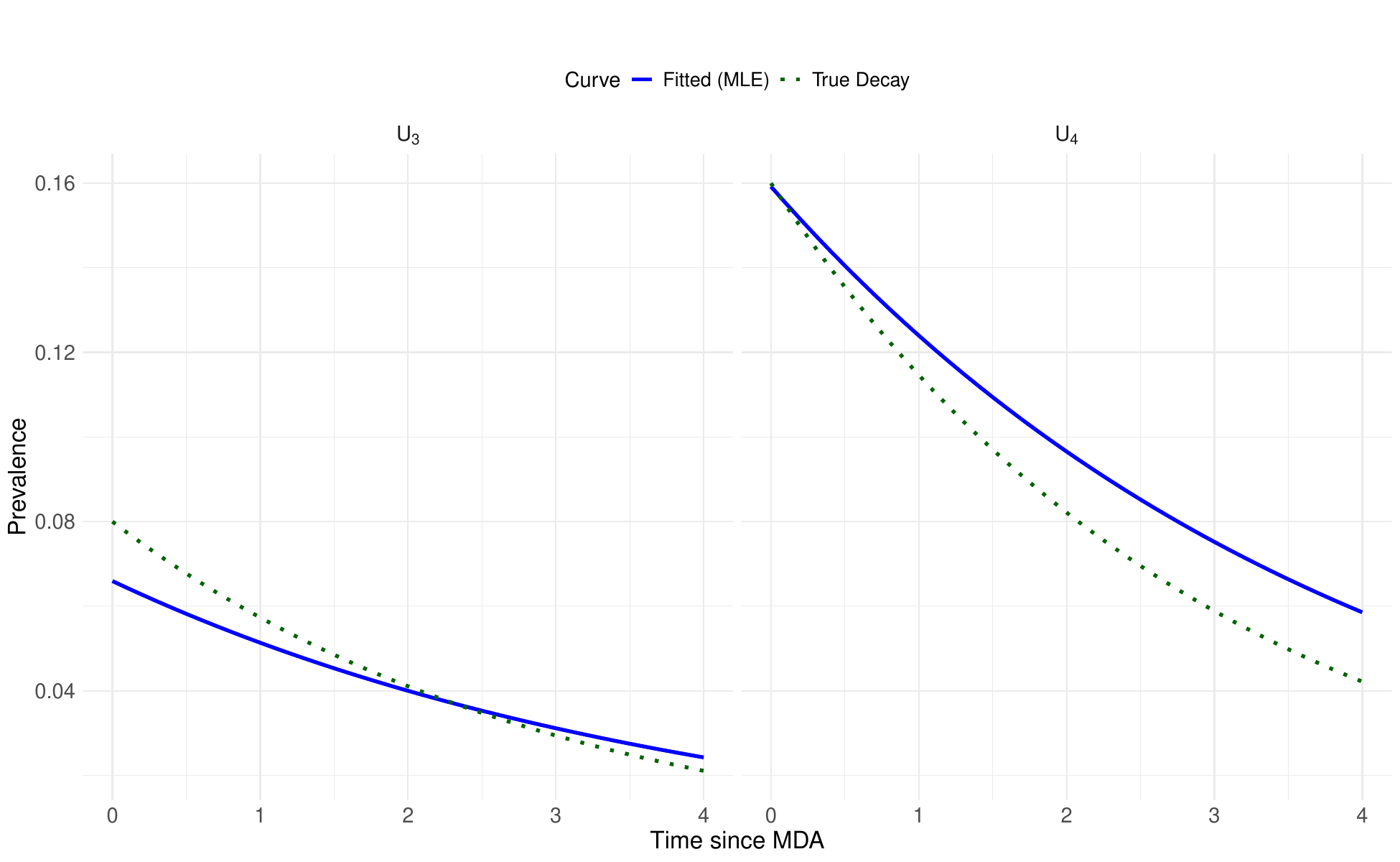}
    \caption{Results from a simulated data-set using the model outlined in Section \ref{subsec:survey_conf}. Estimated and true decay in prevalence following a single round of MDA administered at $t = 0$, plotted as a function of time since treatment. The fitted curves (solid blue) are based on maximum likelihood estimation under the decay-adjusted model, while the true decay curves (dotted green) represent the data-generating process. For more details, see the main text in Section \ref{subsec:survey_conf}.}
    \label{fig:sim_example}
\end{figure}

This extreme example illustrates the fundamental problem with using the cumulative number of MDA rounds as an unconstrained covariate. When survey design induces perfect collinearity between covariates of interest and spatial or temporal structure, regression coefficients become non-identifiable, and standard modelling approaches fail. Even in less extreme settings, similar biases can occur, leading to seemingly paradoxical associations, such as increased MDA coverage appearing to be associated with higher prevalence. This phenomenon is also known as \textit{confounding by indication}, a well recognised limitation of observational studies in which interventions are preferentially implemented in populations with the greatest underlying risk \citep{Psaty1999}.

In this example, a viable alternative is to fit the model from which the data were generated, as specified in equation~\eqref{eq:model_sim_example}. Unlike the standard generalized linear model in~\eqref{eq:glm_naive}, the decay-adjusted formulation remains identifiable even under the spatial confounding induced by the survey design. Figure~\ref{fig:sim_example} displays both the true and estimated effect of MDA on prevalence as a function of time since the single MDA round administered at $t = 0$ using a single simulated data-set where we set $\gamma=3$. This approach yields plausible estimates that correctly capture the underlying trend of decreasing prevalence with increasing time since treatment. Hence, this simple example demonstrates that incorporating a more epidemiologically plausible structure  into the model,
in this case through an exponential decay, enables valid inference even when standard covariate-based approaches fail due to collinearity. 

In the next section, we extend this principle to geostatistical models, where the same structural constraint can be used to overcome similar identifiability issues and improve both robustness and interpretability of the fitted models.

\section{A decay-adjusted spatio-temporal (DAST) model}\label{sec:methodology}
\label{sec:methods}
\begin{figure}
    \centering
    \includegraphics[scale=0.7]{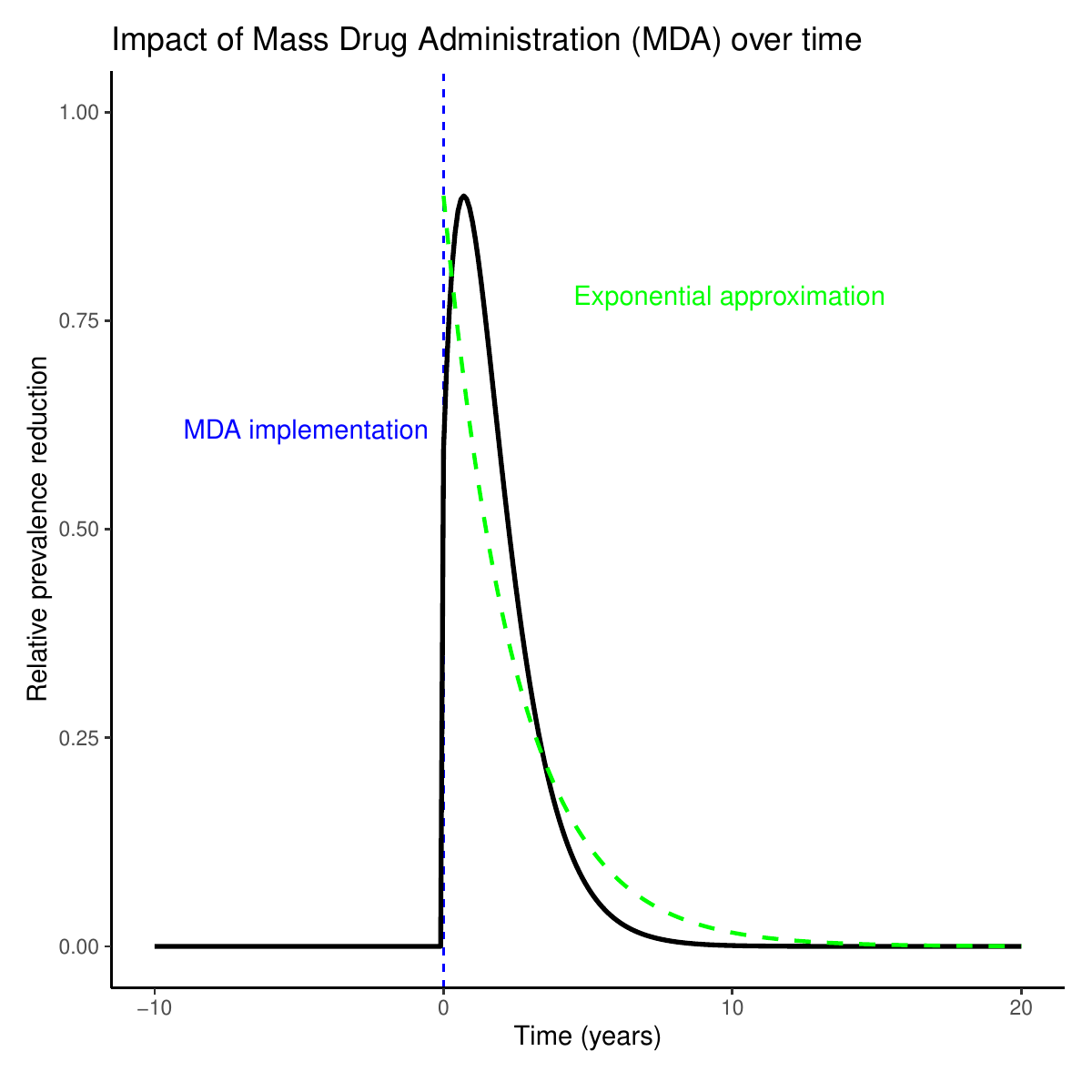}
\caption{Illustration of the expected effect of one round of mass drug administration on disease prevalence over time. The solid black line represents a stylised impact function representing the relative reduction in prevalence due to the impact of treatment. The dashed green line shows the decay approximation used in our proposed decay-adjusted spatio-temporal geostatistical model. The vertical dashed blue line indicates the time of MDA implementation.}    \label{fig:mda_impact}
\end{figure}

\subsection{Model formulation} \label{model}

Figure~\ref{fig:mda_impact} illustrates a stylised representation of the temporal effect of a single round of MDA on disease prevalence. Prior to MDA, prevalence is assumed to remain relatively stable, with the dashed blue line indicating the time of MDA implementation. Following treatment, prevalence declines sharply to a minimum, representing the peak effect, and then gradually increases over time as individuals become reinfected. This rebound is captured by the solid black line. The dashed green curve overlays our proposed approximation of this effect, which is incorporated into a geostatistical framework that accounts for the decay in impact over time.

In our model development, we make the following set of assumptions.

\begin{itemize}
  \item[\textbf{A1}] \textit{Baseline equilibrium.} In the absence of MDA, disease prevalence is assumed to be stationary in time, at any given location $x$, fluctuating around a constant mean, $p(x)$.
  
  \item[\textbf{A2}] \textit{Transient impact.} Each round of MDA induces an immediate relative reduction in prevalence, quickly reaching a peak and then waning progressively if no subsequent rounds are administered.
  
  \item[\textbf{A3}] \textit{Proportional efficacy.} The effect of MDA acts multiplicatively on the underlying prevalence, such that the absolute reduction depends on the pre-intervention prevalence level but the relative reduction is independent of it.
  
  \item[\textbf{A4}] \textit{Survey timing.} When the reported times of the survey and MDA coincide, it is assumed that the survey was conducted immediately before the rollout of MDA.
\end{itemize}

These assumptions provide a parsimonious yet interpretable framework for modelling MDA impact. A1 is a reasonable approximation for many endemic settings where, in the absence of additional interventions, prevalence tends to fluctuate around a stable average over short to medium time scales. A2 reflects a well-documented epidemiological pattern in which prevalence declines sharply following MDA and then gradually returns to baseline as reinfection occurs, behaviour that our flexible decay function is designed to capture. A3 assumes that the relative effect of treatment does not depend on the underlying prevalence level. This may not hold in all settings, particularly those with very high transmission, but is a reasonable simplification in most circumstances. Finally, A4 reflects the level of temporal information typically available for these datasets, where only the year of the survey and the year of MDA implementation are reported. Under this constraint, it is assumed that surveys conducted in the same year as an MDA round take place immediately before its rollout. This assumption is also consistent with programmatic guidance, whereby impact surveys are typically conducted just before MDA implementation and then again 6 to 12 months after five or six consecutive MDA rounds. Under this framework, the observed prevalence in the year following MDA is assumed to reflect the early phase of the decay. Consequently, our modelling focuses on capturing this decay in prevalence rather than reconstructing the full trajectory of the MDA effect.

We now introduce a geostatistical model that satisfies assumptions A1 to A4, while explicitly accounting for the timing and coverage of MDA rounds.
For $i = 1, \ldots, n$, let $Y_i$ denote the number of positive individuals out of $m_i$ tested at location $x_i$ and time $t_i$. We assume that
\begin{equation}
Y_i \mid P(x_i, t_i) \sim \text{Bin}(m_i, P(x_i, t_i)),
\label{eq:binomial}
\end{equation}
where $P(x_i, t_i)$ denotes the observed prevalence at location $x_i$ and time $t_i$, after accounting for the effect of past MDA rounds.

To model the effect of MDA, we assume that the observed prevalence is the product of a baseline prevalence $P^*(x_i)$, the prevalence that would have been observed in the absence of treatment, and a multiplicative decay term that captures the cumulative impact of all previous MDA rounds:
\begin{equation}
P(x, t) = P^*(x) \prod_{j \: : \: u_j < t} \left[1 - f(t - u_j)\right]^{\mathbb{I}(x, u_j)},
\label{eq-prev}
\end{equation}
where the terms $u_j$ and $\mathbb{I}(x, u_j)$ are defined as in equation~\eqref{eq:cum_mda}. The function $f(v)$ quantifies the proportionate reduction in prevalence attributable to an MDA round administered $v$ units of time earlier. We model the log-odds of the baseline prevalence $P^*(x_i)$ as a function of spatially referenced covariates and a latent Gaussian process $S(x)$:
\begin{equation}
\log\left[\frac{P^*(x_i)}{1 - P^*(x_i)}\right] = d(x_i)^\top \beta + S(x_i),
\label{eq:endemic}
\end{equation}
where $d(x_i)$ is a vector of covariates with associated regression coefficients $\beta$, and $S(x)$ is a zero-mean stationary Gaussian process with variance $\sigma^2$ and Matérn correlation function, as defined in Section~\ref{sec:current_geo}.

While the true temporal pattern of post-MDA impact may deviate from a strictly monotonic decline (see Figure \ref{fig:mda_impact}) the coarse temporal resolution of typical NTD survey data does not allow us to reliably estimate a more complex functional form. We therefore specify the $f(v)$ function using an exponential decay formulation as a pragmatic and parsimonious approximation. Hence,
\begin{equation}
f(v) = \alpha \exp( -v/\gamma) , \quad v > 0,
\label{eq:MDA}
\end{equation}
where $\alpha \in [0, 1]$ denotes the immediate post-treatment effect and $\gamma > 0$ controls the scale of the decay over time.

We stress that this parametric form is not intended to capture the full biological or programmatic complexity of MDA effects, but rather to provide a statistically identifiable and interpretable approximation suited to the limitations of the data. The shape of the true MDA impact function illustrated by the solid line in Figure~\ref{fig:mda_impact} reflects the expected biological dynamics of MDA at the population level. Following treatment rollout, prevalence does not decline instantaneously but reaches a minimum some time afterwards due to several reasons. For example, logistical constraints mean that different communities within an implementation unit are reached at different times, so the full effect of a treatment campaign accumulates gradually across the population. Furthermore, even after treatment, individuals remain exposed to transmission and will progressively become reinfected, partially counteracting the effect of MDA. The net result is a curve that peaks some time after the nominal rollout date before declining gradually. Estimating the timing and height of this peak is not feasible with the data typically available in NTD programmes, where surveys are conducted at most once per year and rarely capture the period immediately following treatment. For this reason, we adopt the pragmatic assumption A4 that surveys coinciding with MDA are conducted immediately before rollout, so that all observed post-MDA prevalence reflects the decay phase rather than the rise-to-peak phase. Under this assumption, the exponential decay in equation~\eqref{eq:MDA} provides a parsimonious and interpretable approximation to the waning of the MDA effect, with $\alpha$ capturing the magnitude of the reduction at the point of first post-MDA observation and $\gamma$ controlling the rate at which this reduction diminishes over time. The exponential decay may not capture all the possible shapes of the true MDA impact functions, rather, we argue that this functional form is well suited to the temporal resolution of available survey data, provides clearly interpretable parameters, and represents one of the simplest choices that captures the key features of the post-MDA trajectory observable from routine programme data.

In settings where assumption A1 does not hold, for example due to movement of people over time, or concurrent interventions for other diseases, the model can be extended by allowing the baseline prevalence to vary in time, replacing $ P^*(x) $ with $ P^*(x, t) $. This would involve introducing a spatio-temporal Gaussian process $ S(x, t) $ in place of $ S(x) $, and using time-varying covariates $ d(x, t) $ to help explain additional variation in prevalence that is not attributable to MDA. However, for the sake of parsimony and to avoid issues of model identifiability, one may alternatively retain the purely spatial process $ S(x) $ and rely on suitably chosen spatio-temporal covariates to account for temporal fluctuations in the underlying risk surface.

A special case of the impact function arises when $ f(v) = \alpha $, i.e., a constant non-waning effect that does not satisfy assumption A2. In this case, the product term in equation~\eqref{eq-prev} simplifies to
\begin{equation}
\label{eq:p_xt}
P(x, t) = P^*(x) \prod_{j \: : \: u_j < t} (1 - \alpha)^{\mathbb{I}(x, u_j)} = P^*(x) \, (1 - \alpha)^{c(x,t)},    
\end{equation}
where $ c(x,t)$ denotes the cumulative number of effective MDA rounds received at location $ x $ prior to time $ t $, as previously defined in \eqref{eq:cum_mda}. The model thus reduces to a simple exponential attenuation of prevalence with respect to the number of rounds. While such an assumption is not epidemiologically realistic, since the effect of a single round is known to diminish over time, it may be useful in settings where only a small number of repeated cross-sectional surveys are available, making it difficult to identify a more flexible decay function. In such situations, the use of a constant $ f(v) = \alpha $ provides a pragmatic simplification. In particular, this also provides a theoretical basis for approaches that treat the cumulative number of rounds of MDA as a logit-linear predictor of prevalence.

\subsection{Inference}
\label{subsec:inference}

We adopt a Monte Carlo maximum-likelihood (MCML) approach \citep{Geyer1992-gd,christensen2004, christensen2006} to carry out inference for the DAST model. Throughout, we assume assumptions A1 to A4 hold, but the framework can be adapted to accommodate violations of these assumptions.

Recall from Section~\ref{model} that, conditional on the Gaussian process $S(x)$, the observed counts $Y_i$ at locations $x_i$ and times $t_i$, for $i = 1, \dots, n$, are independent binomial random variables:
\begin{equation}
Y_i \mid S(x_i) \sim \text{Bin}\left(m_i, P(x_i, t_i)\right),
\end{equation}
where $P(x_i, t_i)$ is the observed prevalence adjusted for MDA rounds, as specified in equation~\eqref{eq-prev}, and depends explicitly on the latent Gaussian process $S(x_i)$ through the baseline prevalence $P^*(x_i)$, given by equation~\eqref{eq:endemic}.

Let $\boldsymbol\psi = (\boldsymbol\beta, \boldsymbol\theta)$ denote the vector of model parameters, with $\boldsymbol\beta$ representing regression coefficients associated with covariates and $\boldsymbol\theta$  governing the spatial covariance of $S(x)$. The marginal likelihood is obtained by integrating out the latent Gaussian process:
\begin{equation}
L(\boldsymbol\psi) = \int_{\mathbb R^n}
\phi_n(\mathbf{S}; \mathbf{0}, \Sigma(\boldsymbol\theta))
\prod_{i=1}^n \text{Bin}\left(y_i ; m_i, P(x_i, t_i; S(x_i), \boldsymbol\beta)\right) \,\mathrm d \mathbf{S},
\label{eq:likelihood-marginal}
\end{equation}
where $\mathbf{S} = (S(x_1), \dots, S(x_n))^\top$ is the vector of latent Gaussian variables, $\phi_n(\cdot; \mathbf{0}$, while $\Sigma(\boldsymbol\theta))$ denotes the $n$-dimensional Gaussian density with zero mean and covariance matrix $\Sigma(\boldsymbol\theta)$ structured according to the Matérn correlation function.

Since the integral in equation~\eqref{eq:likelihood-marginal} is analytically intractable, we approximate it using Monte Carlo methods. Specifically, we employ importance sampling. Following \citet{geyer1994}, we choose a reference parameter vector $\boldsymbol\psi^{(0)} = (\boldsymbol\beta^{(0)}, \boldsymbol\theta^{(0)})$, and rewrite the likelihood as an expectation with respect to the conditional distribution of the latent Gaussian process given the observed data:
\begin{equation}
L(\boldsymbol\psi) = \mathbb E_{\mathbf{S} \mid y, \boldsymbol\psi^{(0)}}\left[
\frac{\phi_n(\mathbf{S}; \mathbf{0}, \Sigma(\boldsymbol\theta))}
     {\phi_n(\mathbf{S}; \mathbf{0}, \Sigma(\boldsymbol\theta^{(0)}))}
\prod_{i=1}^n \frac{\text{Bin}\left(y_i ; m_i, P(x_i, t_i; S(x_i), \boldsymbol\beta)\right)}
{\text{Bin}\left(y_i ; m_i, P(x_i, t_i; S(x_i), \boldsymbol\beta^{(0)})\right)}
\right].
\label{eq:importance-sampling}
\end{equation}

To approximate equation~\eqref{eq:importance-sampling}, we draw $m$ independent samples $\mathbf{S}^{(1)}, \dots, \mathbf{S}^{(m)}$ from the conditional distribution $p(\mathbf{S} \mid y, \boldsymbol\psi^{(0)})$. The Monte Carlo approximation of the likelihood is thus given by:
\begin{equation}
L_m(\boldsymbol\psi) = \frac{1}{m}\sum_{l=1}^m \frac{\phi_n(\mathbf{S}^{(l)}; \mathbf{0}, \Sigma(\boldsymbol\theta))}{\phi_n(\mathbf{S}^{(l)}; \mathbf{0}, \Sigma(\boldsymbol\theta^{(0)}))}
\prod_{i=1}^n \frac{\text{Bin}\left(y_i ; m_i, P(x_i, t_i; S^{(l)}(x_i), \boldsymbol\beta)\right)}
{\text{Bin}\left(y_i ; m_i, P(x_i, t_i; S^{(l)}(x_i), \boldsymbol\beta^{(0)})\right)}.
\label{eq:likelihood-mcml}
\end{equation}

We then maximise the approximate log-likelihood $\ell_m(\boldsymbol\psi) = \log L_m(\boldsymbol\psi)$ using quasi-Newton optimisation methods. To enhance numerical stability, gradients with respect to $(\boldsymbol\beta, \log \boldsymbol\theta)$ are typically used during optimisation.

To generate the Monte Carlo samples from the conditional distribution $\mathbf{S} \mid y, \boldsymbol\psi^{(0)}$, we apply the Laplace-based sampling method described in \citet{christensen2004}. This approach begins by locating the posterior mode of $\mathbf{S}$ under the fixed reference parameters. Around this mode, the posterior distribution is approximated by a multivariate Gaussian with covariance matrix given by the negative inverse of the Hessian of the log-posterior. Samples are then drawn using a Metropolis-adjusted Langevin algorithm (MALA) applied to a reparametrised version of the latent field, scaled according to this local curvature. This yields efficient exploration of the high-dimensional space even in models with strong posterior correlation structures. The proposal step size is tuned to achieve an optimal acceptance rate, typically around 45\% as recommended in \citet{roberts2001}.

The procedure is iterated by updating the reference parameter values to the most recent maximiser, re-sampling the latent field, and re-optimising the approximate likelihood until convergence is achieved. Convergence is assessed through the stability of the log-likelihood estimates and the parameter updates. Standard errors for the final estimates are obtained from the observed information matrix, approximated by the negative Hessian of $\log L_{m}(\boldsymbol{\psi})$ at the maximum.

Both pixel-level and area-level predictions are obtained using plug-in estimates, whereby the model parameters $\boldsymbol{\psi}$ are fixed at their maximum likelihood estimates $\hat{\boldsymbol{\psi}}$. To provide uncertainty summaries for the model parameters, we adopt a parametric bootstrap approach that avoids reliance on asymptotic Gaussian approximations. A key benefit of the proposed implementation is that it provides a computationally efficient data-driven framework for inference. Unlike full Bayesian methods,  the use of MCMC is required only for sampling from the distribution of the latent variable $\mathbf{S}$ given the data while fixing the parameter values for $\boldsymbol{\psi}$. This approach results in good mixing within a relatively small number of iterations and faster computation of the predictive summaries for prevalence. 

A limitation of the plug-in approach is that predictions do not propagate parameter uncertainty, which may lead to
underestimation of predictive uncertainty. One approach to address this within the MCML framework is to use a parametric bootstrap, whereby new datasets are
repeatedly simulated from the fitted model and the variation in the resulting parameter estimates is used to obtain corrected predictive distributions; we
refer the reader to \citet{giorgi2018} for a detailed treatment in the context of geostatistical spatio-temporal models. 

\subsection{Penalisation for stabilising estimation of $\alpha$}
\label{sec:penal_alpha}

Because $\alpha$ represents the impact on disease prevalence immediately after MDA has been administered, it may be difficult to estimate this parameter due to the coarse temporal resolution of NTD survey data. In our applications of the DAST model, we have found that the parameter $\alpha$ is often the most challenging to estimate reliably, particularly in settings with limited spatial overlap between sampled areas across time points.
 In such cases, the likelihood function for $\alpha$ tends to be relatively flat, and when the true value is near $1$, maximum likelihood estimation can yield values extremely close to the boundary. Hence, to mitigate instability in the estimation of $\alpha$, we propose to augment the log-likelihood with a general penalisation term, yielding
\begin{equation}
\log L_{\text{pen}}(\boldsymbol\psi) = \log L(\boldsymbol\psi) - g_{\text{pen}}(\alpha),
\end{equation}
where $g_{\text{pen}}(\alpha)$ is a smooth function designed to discourage implausible or weakly identified values of $\alpha$, particularly near the boundaries of the unit interval. Various forms of penalisation can be considered, including Jeffreys-type penalties or bias-reducing adjustments such as those introduced by \citet{firth1993bias}; however, some of these can be quite computationally expensive in the context of geostatistical models. In our application, we adopt the specific form
\begin{equation}
g_{\text{pen}}(\alpha) = -\left( \lambda_1 \log(\alpha) + \lambda_2 \log(1 - \alpha) \right),
\end{equation}
 the log-density of a $\text{Beta}(\lambda_1 + 1, \lambda_2 + 1)$ distribution (up to a constant). This implies that maximising the penalised log-likelihood $\log L_{\text{pen}}(\boldsymbol\psi)$ with respect to $\alpha$ is equivalent to
computing the maximum a posteriori estimate of $\alpha$ under a Beta prior with parameters $(\lambda_1+1, \lambda_2+1)$. This connection
with Bayesian inference provides a principled probabilistic interpretation of the penalisation whilst preserving the computational efficiency of likelihood-based inference, and avoids the need to place prior distributions on the remaining
model parameters, for which identifiability is less of a concern.

 This choice also provides an intuitive way to constrain $\alpha$ within a realistic range, while preserving flexibility for the data to inform the estimate when sufficient information is available. Constraining the range can be problematic if the range is chosen arbitrarily, as it may unduly bias the inference. For this reason, in this paper we use the penalty solely to discourage extreme estimates too close to the boundaries, rather than to impose a strict restriction on the admissible range of $\alpha$. For example, using values of $(\lambda_1, \lambda_2)$ that favour $\alpha$ in the interval $[0.05, 0.95]$ reflects realistic expectations of MDA impact derived from programmatic experience in a variety of settings. This choice does not impose strong prior information but helps to stabilise inference in data-sparse or poorly identified scenarios. At the same time, by keeping the penalisation relatively light, we aim to retain the ability to use the estimated impact function to draw meaningful inferences on the effect of MDA. One could alternatively adopt a fully Bayesian approach, placing prior distributions on all model parameters. However, we prefer the penalised likelihood formulation used here, as it allows us to selectively regularise only $\alpha$, without introducing unnecessary prior structure on other parameters for which prior information is typically lacking. Alternative strategies for setting $(\lambda_1, \lambda_2)$, such as cross-validation or empirical Bayes tuning, may also be considered depending on the application context and the available data.

This form of penalisation can introduce bias in the estimation of $\alpha$, particularly when the values of $\lambda_1$ and $\lambda_2$ are large, in which case the effect of the penalisation can overwhelm the contribution of the data for relatively small sample sizes. However, in many practical applications, precise inference on $\alpha$ itself is not the primary objective. Instead, the focus lies on accurately capturing the cumulative impact of MDA through the decay function $f(v)$ and generating reliable prevalence predictions. In the simulation study presented in Section \ref{sec:simulation}, we show that penalisation of $\alpha$ serves as a stabilising device with minimal consequence for the overall inferential goals of the analysis.

We do not apply the penalisation by default. Instead, our approach is to first fit the model without any penalisation and assess the identifiability of $\alpha$ using standard diagnostics, such as the shape of the profile likelihood and the stability of the optimisation procedure. The penalty is introduced only if the unpenalised model displays signs of numerical instability or produces implausible boundary estimates for $\alpha$, particularly in scenarios with sparse data or limited temporal overlap in sampling. This two-step strategy ensures that the data are given every opportunity to inform the estimation of $\alpha$, with regularisation applied only as a targeted corrective measure to enable stable and interpretable inference when needed.

\subsection{Predictive targets in the context of neglected tropical disease control}
\label{subsec:prediction}

Under the DAST modelling framework, the primary predictive target is area-level prevalence, defined for each implementation unit (IU) $U_k$, with $k = 1, \ldots, K$, at a given time point $t$. This target is formally expressed as the spatial average of the prevalence over the entire IU:
\begin{equation}
\label{eq:ev_prev}
\mathcal{P}_{t}(U_k) = \frac{1}{|U_k|} \int_{U_k} P(x,t) \, \mathrm{d}x,
\end{equation}
where $|U_k|$ denotes the area of the IU. Such area-level averages play a crucial role in public health by directly informing programmatic decisions, with interventions typically triggered when $\mathcal{P}_{t}(U_k)$ exceeds established policy thresholds. Alternative definitions of the predictive target may also be appropriate depending on the application. For example, one may incorporate a weighting function $w(x)$, such as population density, to reflect the distribution of the at-risk population:
\begin{equation}
\tilde{\mathcal{P}}_{t}(U_k) = \frac{1}{\int_{U_k} w(x) \, \mathrm{d}x} \int_{U_k} w(x) P(x,t) \, \mathrm{d}x.
\end{equation}
This population-weighted prevalence estimate provides a standardised measure that accounts for spatial heterogeneity in population distribution by assigning greater weight to more densely populated areas. In settings where the locations of all communities within an IU are known, the integral in \eqref{eq:ev_prev} can be replaced by a discrete sum over community-level predictions:
\begin{equation}
\mathcal{P}^*_{t}(U_k) = \frac{1}{n_k} \sum_{j=1}^{n_k} P(x_j, t),
\end{equation}
where $x_1, \ldots, x_{n_k}$ are the coordinates of the $n_k$ communities in $U_k$. Although this approach avoids the need for spatial integration, it is less commonly used in practice, as detailed community-level location data are seldom available.

A central challenge in assessing the predictive performance of geostatistical models for this areal-level target arises because $\mathcal{P}_{t}(U_k)$ is a functional of the latent spatial process $P(x,t)$, encompassing every location within the IU. Although cross-validation techniques effectively evaluate calibration and sharpness at specific locations, they cannot adequately assess accuracy at the IU level. Therefore, we rely on simulation studies to examine potential biases and validate the reliability of uncertainty quantification, as described in Section~\ref{sec:simulation}.

The importance of accurately estimating IU-level prevalence is underscored by specific public health examples where defined prevalence thresholds guide intervention strategies. For instance, in the context of trachoma, WHO defines elimination as a public health problem based on two IU-level prevalence thresholds \citep{WHO1996GET}: less than 0.2\% prevalence of trachomatous trichiasis unknown to the health system among adults aged 15 years or older; and less than 5\% prevalence of trachomatous inflammation–follicular among children aged 1–9 years. Both criteria must be simultaneously met in each formerly endemic IU, typically an administrative unit with a population between 100{,}000 and 250{,}000. An example of a model-based geostatistical application to evaluate these thresholds is provided in \citet{sasanami2023}.

Similarly, for soil-transmitted helminths (STH), WHO guidelines define IU-level prevalence categories at baseline as low ($<20\%$), moderate ($\geq20\%~ - <50\%$), or high ($\geq 50\%$). These categories respectively correspond to recommendations of no preventive chemotherapy, annual treatment, or biannual treatment. During subsequent impact assessments, the prevalence classification expands to five categories, reflecting more detailed shifts in infection intensity over time. These prevalence categories critically shape operational planning and survey design, as illustrated in \citet{World-Health-Organization2024-kn}.

For schistosomiasis, the WHO classifies IUs according to the infection prevalence in school-aged children as below  or above 10\%  \citep{World-Health-Organization2024-kn}. In IUs with a prevalence below $10\%$, preventive chemotherapy is not routinely required, and a clinical case management approach is recommended. In IUs with a prevalence of $10\%$ or higher, PC should be distributed once per year to the entire population aged two years and above for five consecutive years, followed by an impact assessment. If the subsequent prevalence is found to be below $10\%$, the frequency of PC may be reduced or discontinued based on historical evidence and local transmission factors, otherwise, annual PC should continue.

\section{Simulation studies}
\label{sec:simulation}

\subsection{Assessment of the predictive performance under spatially misaligned sampling}

We carry out a simulation study to achieve the following objectives:
\begin{itemize}
    \item[(i)] to quantify the impact of spatially misaligned sampling over time on areal-level predictive inferences;
    \item[(ii)] to quantify the impact of the proposed penalization for the parameter $\alpha$ of the MDA effect on areal-level predictive inferences.
\end{itemize}

To this end, we consider a hypothetical study region defined on the unit square $[0,1] \times [0,1]$, partitioned into four IUs, denoted by $U_k$ for $k = 1,\ldots,4$, as shown in Figure~\ref{fig-overview}. Data are simulated at four time points ($t = 0, 1, 2, 3$), with 100 locations sampled at each time point, for a total sample size of 400 locations. The model is fitted to data from these four time points and subsequently used to predict prevalence at $t=0, 1, 2, 3$ and to forecast prevalence at a fifth time point, $t = 4$. We assume that MDA is implemented at all locations at each time point $t = 0, \ldots, 3$. The data generated at $t = 0$ thus correspond to the baseline, as they are assumed to have been collected prior to the rollout of MDA.

We compare two sampling scenarios. In Scenario~1, each annual survey selects 100 locations at random across all four IUs, ensuring full spatial and temporal coverage. This represents an idealised setting with well-balanced and temporally aligned survey effort. In contrast, Scenario~2 introduces spatial–temporal misalignment by sampling all 100 observations at each time point from a single, randomly chosen IU. This design reduces spatial overlap across time, limiting the model’s ability to recover temporal dynamics from the data, within any given IU. Hence, we can interpret Scenario~2 as a lower bound on expected model performance, reflecting more challenging operational conditions where resources and logistics constrain the spatial coverage of the survey.

In each simulation replicate, the model is fitted twice: once with penalisation on $\alpha$ and once without. When penalisation is applied, it is implemented by setting $\lambda_1 = \lambda_2 \approx 0.35$, corresponding to a Beta distribution on $\alpha$ whose 2.5th and 97.5th percentiles are 0.05 and 0.95, respectively.

The baseline logit-prevalence surface in the absence of MDA is specified as:
\begin{equation}
\log\left\{ \frac{P^*(x)}{1 - P^*(x)} \right\} = \sum_{k=1}^4 \beta_k \, \mathbb{I}_k(x) + S(x),
\label{eq:baseline}
\end{equation}
where $\beta_k$ are fixed effects set to $-1$, $-1/2$, $1/2$, and $1$ for $U_1$ through $U_4$, respectively; $\mathbb{I}_k(x)$ is an indicator for location $x$ belonging to $U_k$; and $S(x)$ is a mean-zero Gaussian process with variance $\sigma^2 = 1$ and exponential correlation function $\rho(u) = \exp(-u / \phi)$, with $\phi = 0.2$. The MDA effect parameters $(\alpha, \gamma) = (0.8, 1/\log(2 \times 0.8))$ which lead to a halving of prevalence one year after a single MDA round according to an exponential decay. Prevalence observations at each location and time point follow a conditional Binomial distribution $Y_i \: | \: P(x_i, t_i) \sim \text{Bin}(m_i, P(x_i, t_i)), \text{with } m_i=50,$ where $P(x,t)$ denotes prevalence adjusted for cumulative MDA rounds as defined in \eqref{eq-prev}. The choice of $m_i = 50$ reflects a typical sample size per location in NTD surveys.

\subsubsection{Evaluation metrics}
\label{subsec:metrics}

Because measures of predictive uncertainty on the prevalence scale depend on the underlying mean prevalence, it is important to use performance metrics that remain interpretable and comparable across time points and regions in the simulation study. We therefore summarise predictive performance using a standardised prevalence scale that removes the dependence of prediction uncertainty on the mean prevalence level.

Let $\widehat{P}_{k,t}^{(r)}$ and $P_{k,t}^{(r)}$ denote the predicted and true IU--level prevalences for IU $U_k$, time $t$, and simulation replicate $r$, and let $\widehat{L}_{k,t}^{(r)}$ and $L_{k,t}^{(r)}$ be their logits. To obtain a measure of predictive uncertainty that is comparable across prevalence levels, we first compute, for each IU--time pair $(k,t)$, the standard deviation of the predicted logits across replicates,
$$S^{L}_{k,t} = {\rm sd}\big(\widehat{L}_{k,t}^{(r)}\big).$$
We then map this spread onto the prevalence scale using the derivative of the inverse--logit function, evaluated at the predicted prevalence,
$$D_{k,t}^{(r)} = \widehat{P}_{k,t}^{(r)}\bigl(1-\widehat{P}_{k,t}^{(r)}\bigr),$$
so that the delta--method approximation to the predictive standard deviation on the prevalence scale is
$$S^{P,(r)}_{k,t} = D_{k,t}^{(r)}\, S^{L}_{k,t}.$$
Finally, we define the standardised prediction error
\begin{equation}
\label{eq:std_pred_err}
    e_{k,t}^{(r)} = \frac{\widehat{P}_{k,t}^{(r)} - P_{k,t}^{(r)}}{S^{P,(r)}_{k,t}},
\end{equation}
which expresses errors in units of the model's own predictive uncertainty. As our interest lies in overall performance rather than IU-specific behaviour, we
average errors across the four IUs within each replicate:
$$\bar{e}_{t}^{(r)}
=
\frac{1}{4}\sum_{k=1}^{4} e_{k,t}^{(r)},
\qquad
\overline{\rm MSE}_{t}^{(r)}
=
\frac{1}{4} \sum_{k=1}^{4} \bigl(e_{k,t}^{(r)}\bigr)^2.$$
Finally, for each time point and for each combination of sampling scenario and
penalisation setting, we summarise performance across simulation replicates using
$${\rm sBias}_t = 
\frac{1}{R}\sum_{r=1}^{R}\bar e_t^{(r)},
\qquad
{\rm sRMSE}_t = 
\sqrt{\frac{1}{R} \sum_{r=1}^{R}\overline{\rm MSE}_t^{(r)}}.$$
\begin{figure}[!ht]
\centering
\includegraphics[width=1\textwidth]{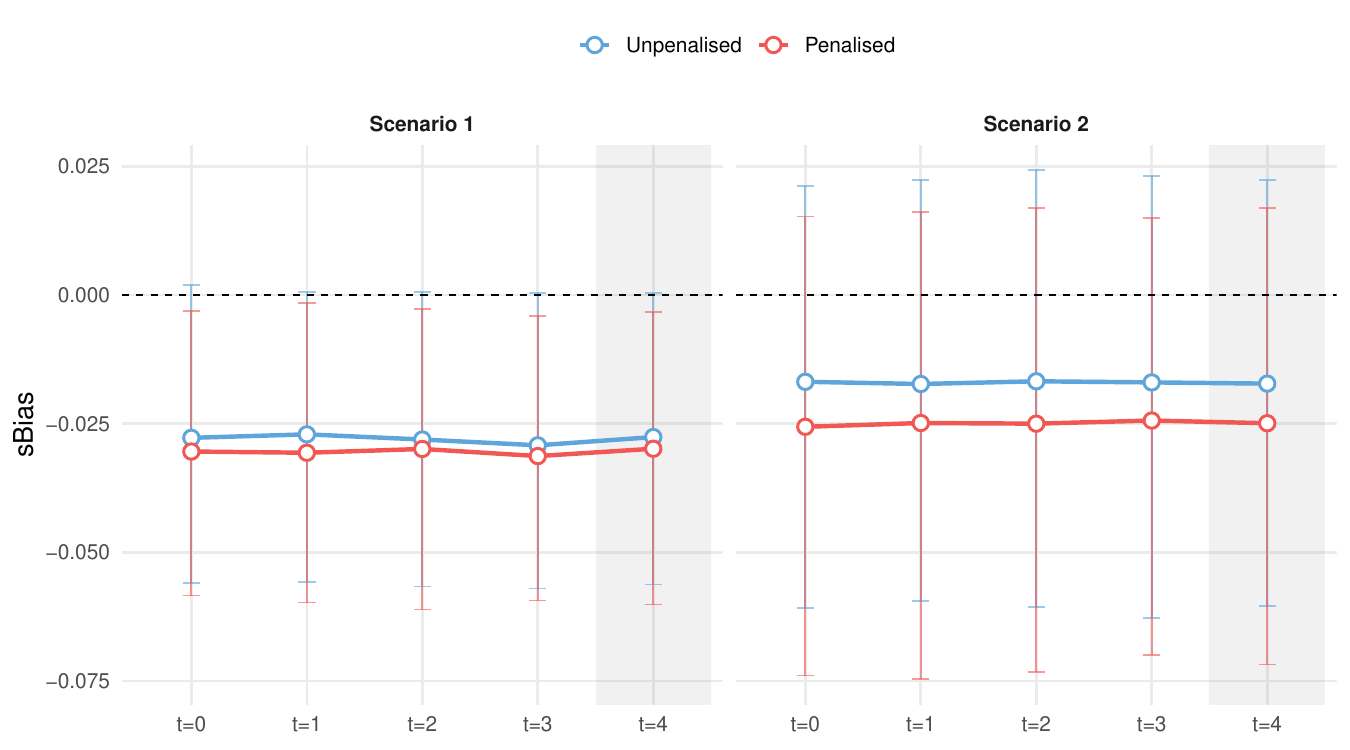}
\caption{Mean standardised bias $\mathrm{sBias}_t$ across simulation replicates, shown
for each sampling scenario, time point, and penalisation setting. The points
represent the mean across replicates (defined in the main text as sBias$_t$), vertical bars indicate the
25th--75th percentile range. The shaded
background marks the forecast time point $t=4$. Unpenalised fits are shown in
blue and penalised fits in red.}
\label{fig:sBias}
\end{figure} 

\begin{figure}[!ht]
\centering
\includegraphics[width=1\textwidth]{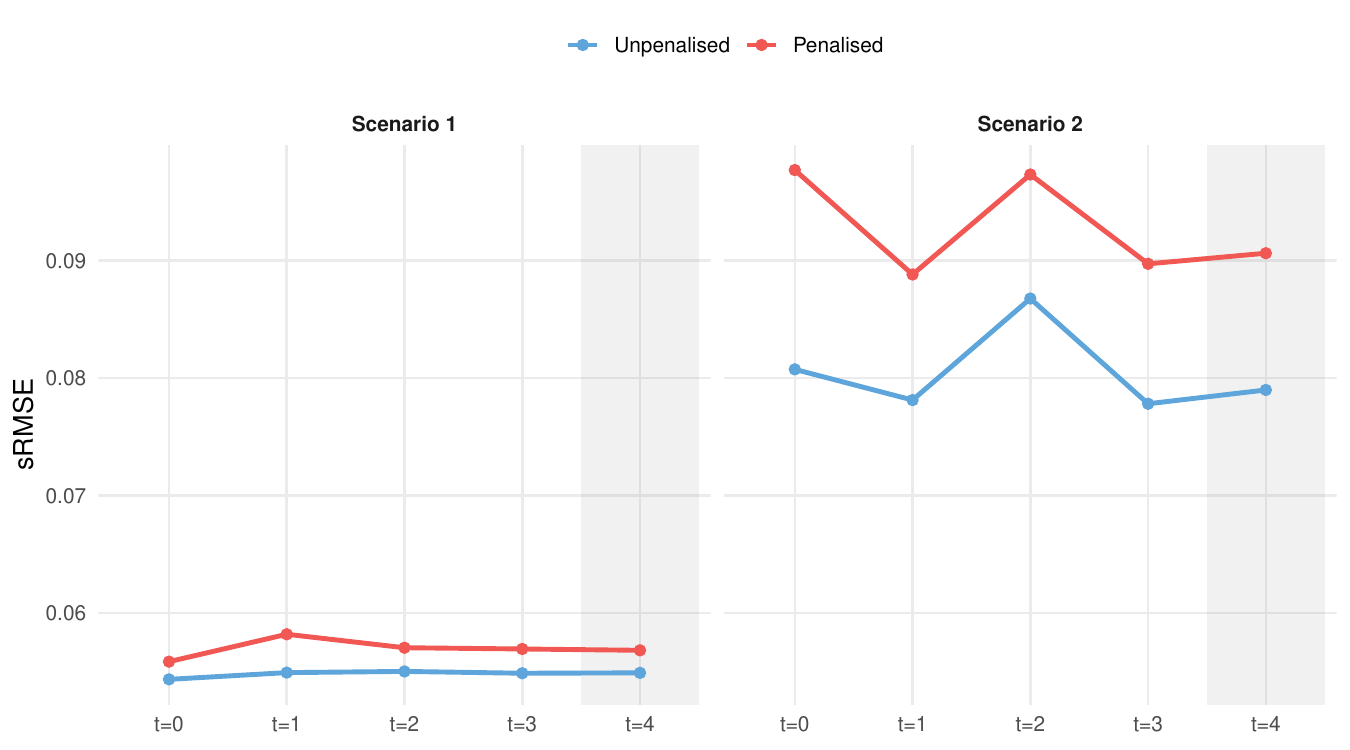}
\caption{Standardised mean square error $\mathrm{sRMSE}_t$ across simulation
replicates for each sampling scenario, time point, and penalisation setting.
Shaded panels highlight the forecast time point
($t=4$). Unpenalised fits are shown in blue and penalised fits in red.}
\label{fig:sRMSE}
\end{figure}

\begin{figure}[!ht]
\centering
\includegraphics[width=1\textwidth]{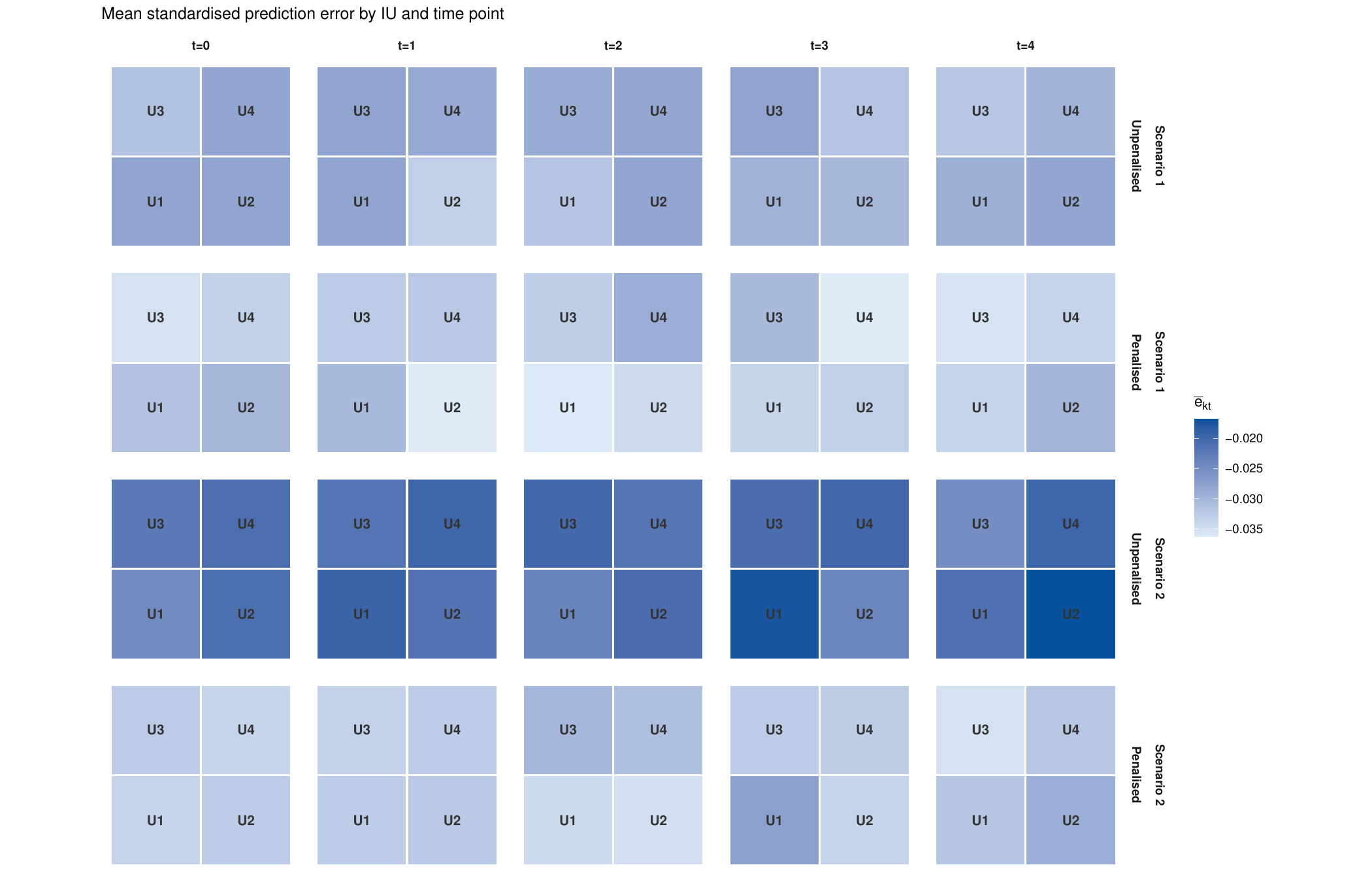}
\caption{Mean standardised prediction error $\bar{e}_{k,t}$ (equation \eqref{eq:std_pred_err}), averaged across
simulation replicates, displayed spatially for each implementation unit $U_k$,
time point, sampling scenario and penalisation setting.}
\label{fig:choro_e}
\end{figure}

\subsubsection{Results}

Figures \ref{fig:sBias}, \ref{fig:sRMSE} and \ref{fig:choro_e} summarise the predictive performance across the 1,000 simulation replicates using the standardised
metrics introduced in Subsection \ref{subsec:metrics}.  For each combination of
sampling scenario, penalisation setting, and time point, the plots display the
median sBias$_t$ with its interquartile range,
and the corresponding sRMSE$_t$.  Both
quantities are defined relative to the model's predictive scale on the prevalence domain and therefore allow comparisons across time points and prevalence levels.

Figure \ref{fig:sBias} shows that systematic deviations are small in all settings.  In Scenario 1, where all four IUs are sampled at every time point, both penalised and unpenalised fits yield nearly identical curves, with
sBias$_t$ consistently around $-0.027$ and narrow interquartile ranges.  This
indicates that, under well-balanced spatial--temporal coverage, the use of a soft penalisation has little effect on the direction or magnitude of systematic prediction error. In Scenario 2, where each year the sampling is confined to a single IU, the variability of sBias$_t$ increases, as expected under reduced temporal alignment.  At the same time, the mean sBias$_t$ values shift slightly upward (closer to zero) for
both penalised and unpenalised fits.  This reflects the denser within-IU
sampling (100 observations per IU per time point, compared with $\approx$25 in Scenario 1), which improves the estimation of IU-level means despite the limited temporal overlap.  In this sparser design, the penalised fit exhibits a somewhat
more pronounced negative bias than the unpenalised fit, but the magnitude of this difference remains small and well within the predictive scale implied by the model.  Crucially, bias remains acceptable for both approaches, and the forecast time point ($t=4$) does not show evidence of systematic drift.

Figure~\ref{fig:sRMSE} displays the corresponding sRMSE$_t$ values.  In
Scenario~1, standardised errors are tightly concentrated across all time points,
with sRMSE$_t$ lying between $0.055$ and $0.06$ for both penalised and
unpenalised fits.  This again confirms that, when the spatial and temporal
structure of the data is rich, penalisation has minimal impact on predictive
accuracy. In Scenario 2, sRMSE$_t$ increases to the range $0.08$--$0.10$, reflecting the
greater predictive uncertainty induced by spatial--temporal misalignment.
Differences between penalised and unpenalised fits become more pronounced in
this setting. The penalised model systematically produces slightly larger standardised errors
at several time points, while the unpenalised model achieves marginally smaller
sRMSE$_t$. Nonetheless, the differences remain modest, and both approaches retain stable behaviour at the forecast time point.

Overall, the model displays good predictive stability across all settings. Systematic bias remains small relative to the model’s predictive scale, and standardised RMSE grows only moderately under spatial--temporal misalignment.  Penalisation has negligible impact under full spatial coverage (Scenario 1) and
introduces only minor differences under the more challenging design (Scenario 2).  These findings indicate that the proposed spatio--temporal model is robust to the choice of penalisation and that survey design, particularly the balance between spatial density and temporal overlap, plays a more central role in determining predictive performance.

\subsection{Assessment of the identifiability of the MDA impact function under confounding by indication}

We carry out a second simulation study with two objectives:
\begin{itemize}
    \item[(i)] to investigate the impact of confounding by indication on inference
    for the MDA impact function parameters $\alpha$ and $\gamma$;
    \item[(ii)] to assess whether penalisation on $\alpha$ can help mitigate the
    identifiability issues that arise under confounding.
\end{itemize}
The study region is defined on the unit square $[0,1]\times[0,1]$, partitioned
into $K = 100$ equal-area IUs arranged on a $10\times 10$ grid,
so that each IU has side length $0.1$. A single round of MDA is administered
at $t=0$ to the top 50\% of IUs by baseline prevalence, representing a scenario in which treatment is preferentially targeted to
higher-burden areas. Data are simulated at $t=0$, when all $K$ IUs are
surveyed (baseline), and at four post-MDA time points $t=1,2,3,4$, when a
random subset of 15\% of IUs is surveyed at each time point. Within each surveyed IU, three locations are sampled uniformly at random,
with $m_i = 50$ per location, giving an expected total of approximately 480 observations per replicate.

Confounding by indication is induced by varying the degree to which post-MDA
surveys are concentrated in high-prevalence IUs. Three scenarios are
considered. Under the neutral scenario $C_0$, the 15\% of IUs surveyed at
each post-MDA time point are selected by simple random sampling, with no
preference for high- or low-burden areas. Under the medium confounding
scenario $C_1$, IUs are selected by weighted sampling with weights
proportional to $0.5 + 0.5r$, where $r \in [0,1]$ denotes the normalised
baseline prevalence rank, so that the highest-burden IU is twice as likely to
be surveyed as the lowest. Under the strong confounding scenario $C_2$,
weights are proportional to $0.25 + 0.75r$, giving a four-fold preference for
high-burden IUs.

The true parameter values are $(\alpha, \gamma) = (0.8,\; 1/\log(2 \times
0.8))$, $\sigma^2 = 1$, $\phi = 0.1$ and $\beta = -1$. The choice $\phi =
0.1$ gives a spatial correlation of approximately $0.37$ between adjacent IU
centroids. To address objective~(ii), each simulation is repeated twice: once
without penalisation on $\alpha$, and once with penalisation using
$\lambda_1 = \lambda_2 = 0.35$, as described in Section~\ref{sec:penal_alpha}.

For each confounding scenario, we report the average point estimate, bias,
empirical standard error, root mean square error, and the 2.5th and 97.5th
percentiles of the sampling distribution of $\hat{\alpha}$ and $\hat{\gamma}$
across replicates. We also report the empirical correlation between $\hat{\alpha}$ and
$\hat{\gamma}$ across replicates as a measure of joint identifiability, with a
strong correlation in either direction indicating that the two parameters are
confounded with one another as a result of a flat likelihood surface.

\subsubsection{Results}

\begin{table}[!h]
\centering
\caption{\label{tab:sim2_combined}Summary of the sampling distribution of $\hat{\alpha}$ and $\hat{\gamma}$ under confounding scenarios $C_0$, $C_1$ and $C_2$ with $K = 100$ implementation units. Results are shown for fits without penalisation on $\alpha$ (UP) and with penalisation $\lambda_1 = \lambda_2 = 0.35$ (P). Mean: average point estimate; Bias: mean signed deviation from true value; SE: empirical standard error; RMSE: root mean square error; 95\% PI: percentile interval of the sampling distribution.}
\centering
\resizebox{\ifdim\width>\linewidth\linewidth\else\width\fi}{!}{
\fontsize{9}{11}\selectfont
\begin{tabular}[t]{llllllll}
\toprule
\multicolumn{1}{c}{ } & \multicolumn{1}{c}{ } & \multicolumn{2}{c}{$C_0$} & \multicolumn{2}{c}{$C_1$} & \multicolumn{2}{c}{$C_2$} \\
\cmidrule(l{3pt}r{3pt}){3-4} \cmidrule(l{3pt}r{3pt}){5-6} \cmidrule(l{3pt}r{3pt}){7-8}
Parameter & Metric & UP & P & UP & P & UP & P\\
\midrule
\addlinespace[0.3em]
\hspace{1em}$\hat{\alpha}$ (true = 0.800) & Mean & 0.820 & 0.789 & 0.819 & 0.797 & 0.823 & 0.801\\
\hspace{1em} & Bias & 0.020 & -0.011 & 0.019 & -0.003 & 0.023 & 0.001\\
\hspace{1em} & SE & 0.124 & 0.104 & 0.112 & 0.094 & 0.106 & 0.091\\
\hspace{1em} & RMSE & 0.126 & 0.105 & 0.114 & 0.094 & 0.109 & 0.091\\
\hspace{1em} & 95\% PI & {}[0.575,\;1.000] & {}[0.563,\;0.954] & {}[0.588,\;1.000] & {}[0.603,\;0.954] & {}[0.618,\;1.000] & {}[0.612,\;0.953]\\
\addlinespace[0.3em]
\hspace{1em}$\hat{\gamma}$ (true = 2.128) & Mean & 1.847 & 1.912 & 1.838 & 1.877 & 1.813 & 1.859\\
\hspace{1em} & Bias & -0.280 & -0.215 & -0.289 & -0.251 & -0.314 & -0.269\\
\hspace{1em} & SE & 0.396 & 0.394 & 0.353 & 0.318 & 0.320 & 0.320\\
\hspace{1em} & RMSE & 0.485 & 0.449 & 0.456 & 0.405 & 0.449 & 0.418\\
\hspace{1em} & 95\% PI & {}[1.308,\;2.802] & {}[1.397,\;2.833] & {}[1.312,\;2.666] & {}[1.419,\;2.631] & {}[1.314,\;2.536] & {}[1.406,\;2.573]\\
\bottomrule
\end{tabular}}
\end{table}

\begin{figure}
    \centering
    \includegraphics[width=1\linewidth]{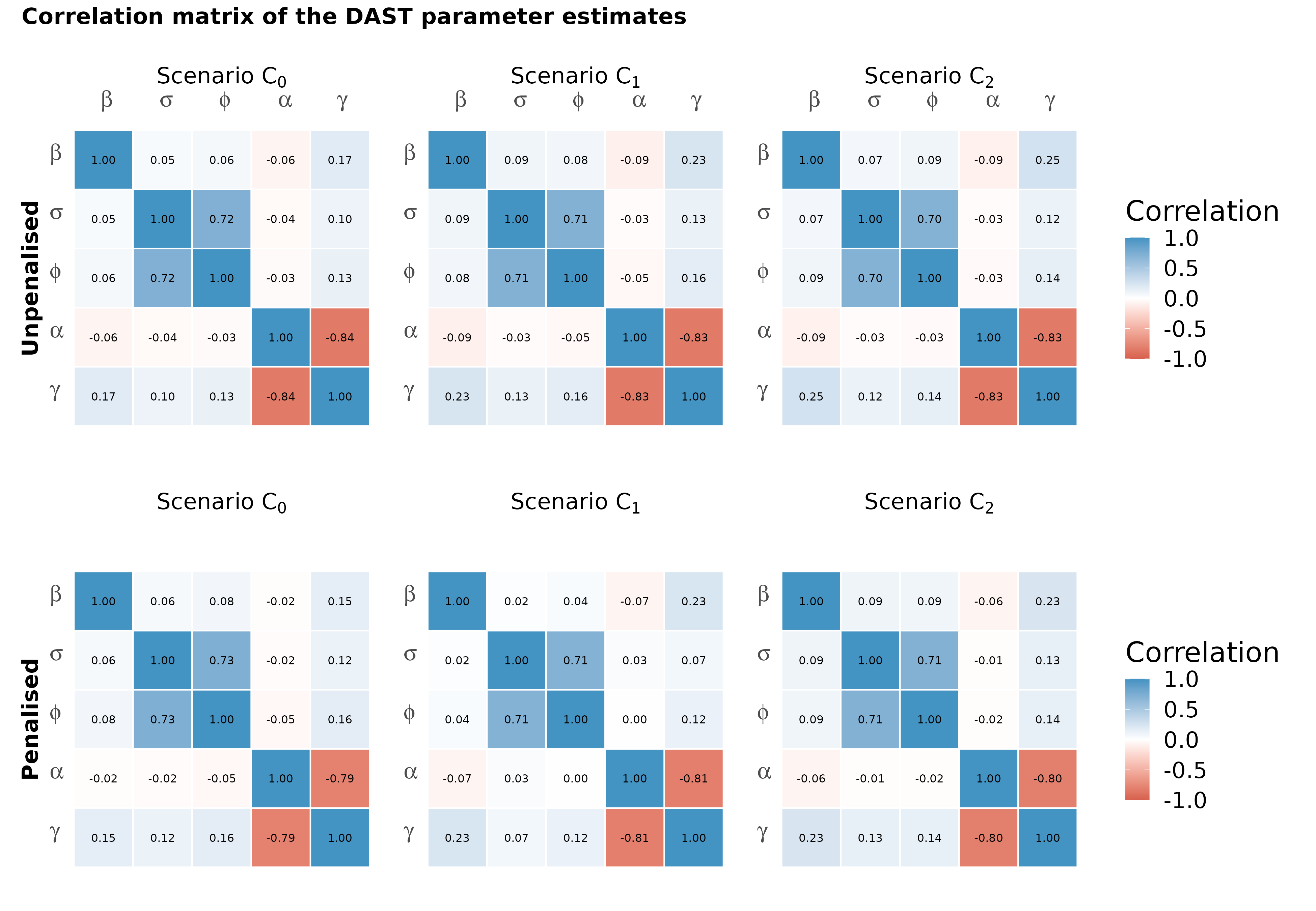}
    \caption{Correlation matrix of the DAST parameter estimates
    ($\hat{\beta}$, $\hat{\sigma}$, $\hat{\phi}$, $\hat{\alpha}$,
    $\hat{\gamma}$) across simulation replicates under confounding scenarios
    $C_0$, $C_1$ and $C_2$ with $K = 100$ implementation units. The upper row shows results without penalisation on $\alpha$ and the lower row with penalisation
    ($\lambda_1 = \lambda_2 = 0.35$).}
    \label{fig:corr_mat}
\end{figure}

Table~\ref{tab:sim2_combined} and Figure~\ref{fig:corr_mat} summarise the
estimation performance of the DAST model under the three confounding scenarios,
with and without penalisation on $\alpha$.

For $\hat{\alpha}$, without penalisation the mean estimate is slightly above
the true value of $0.8$ across all scenarios (around $0.820$), and the 97.5th
percentile of the sampling distribution reaches exactly $1.000$, confirming
that a non-trivial fraction of replicates produce estimates at the upper
boundary of the parameter space. Penalisation resolves this issue
entirely, as the 97.5th percentile drops to approximately $0.954$ across all
scenarios, bias becomes negligible (at most $-0.011$), and both SE and RMSE
are reduced by roughly $15\%$. As confounding increases from $C_0$ to $C_2$, estimation of $\hat{\alpha}$ improves marginally under both penalised and unpenalised fits. However, as we discuss below, this may simply reflect the  negative correlation between
$\hat{\alpha}$ and $\hat{\gamma}$, whereby a reduction in the bias of one
parameter is accompanied by an increase in the bias of the other.

For $\hat{\gamma}$, without penalisation, the
mean estimates show a bias of approximately $-0.28$ to $-0.31$, and RMSE of
$0.45$--$0.49$. Penalisation reduces bias modestly (to $-0.22$ to $-0.27$)
and improves RMSE slightly across all scenarios. 
However, both in the unpenalised and penalised simulations, bias in
$\hat{\gamma}$ becomes more negative as confounding increases from $C_0$ to
$C_2$. This occurs because post-MDA surveys concentrated in high-prevalence treated
IUs provide limited information on the long-run trajectory of the decay function, which is primarily identified through observations from lower-burden areas at later time points. As this information becomes increasingly scarce
under stronger confounding, the model infers a faster-than-true rebound in prevalence, resulting in systematic underestimation of $\gamma$.

Figure~\ref{fig:corr_mat} shows the correlation matrix of the DAST
parameter estimates across replicates for each confounding scenario, with and
without penalisation. The parameters $\hat{\sigma}^2$ and $\hat{\phi}$
display high positive correlation reflecting the well-documented challenge inherent to the estimation of the spatial covariance parameters. The MDA parameters
$\hat{\alpha}$ and $\hat{\gamma}$ are largely uncorrelated with $\hat{\beta}$,
$\hat{\sigma}^2$ and $\hat{\phi}$, indicating that, as expected, the spatial and intervention
components of the model are identified from largely separate aspects of the
data. A key feature is the strong negative correlation between $\hat{\alpha}$ and
$\hat{\gamma}$, approximately $-0.84$ without penalisation and $-0.79$ with
penalisation, reflecting a ridge in the likelihood surface along which the
magnitude and duration of the MDA effect trade off against one another, so
that the data cannot uniquely determine whether a large immediate reduction
wanes quickly or a more modest reduction persists for longer. This correlation is largely unaffected by the degree
of confounding, remaining stable across $C_0$, $C_1$ and $C_2$ in both rows
of the figure, indicating that the degree of joint identifiability of $(\alpha,
\gamma)$ is an intrinsic feature of the likelihood surface rather than a
consequence of confounding per se.

Overall, these results suggest that $\alpha$ and $\gamma$ are challenging but not impossible to estimate, as performance remains satisfactory across all scenarios, with bias remaining below $15\%$ of the true value for both
parameters. Penalisation on $\alpha$ provides a useful stabilising effect, reducing boundary estimates and modestly improving inference on $\gamma$ as a side effect. More informative penalties, informed by prior programmatic knowledge of plausible ranges for $\alpha$ and $\gamma$, could further alleviate the identifiability challenge by restricting the optimiser to epidemiologically meaningful regions of the parameter space.

\section{Applications}
\label{sec:applications}

\subsection{Case study I: soil transmitted helminths in Kenya}
\label{sec:sth_kenya}

Soil-transmitted helminths are intestinal parasites that infect humans through contact with soil contaminated by fecal matter. The three main species infecting humans are \textit{Ascaris lumbricoides} (roundworm), \textit{Trichuris trichiura} (whipworm), and the hookworms, \textit{Necator americanus} and \textit{Ancylostoma duodenale}. These infections are endemic in many low-resource settings and are associated with anemia, growth retardation, and impaired cognitive development, particularly among school-aged children.

\begin{figure}
    \centering
    \includegraphics[width=1\linewidth]{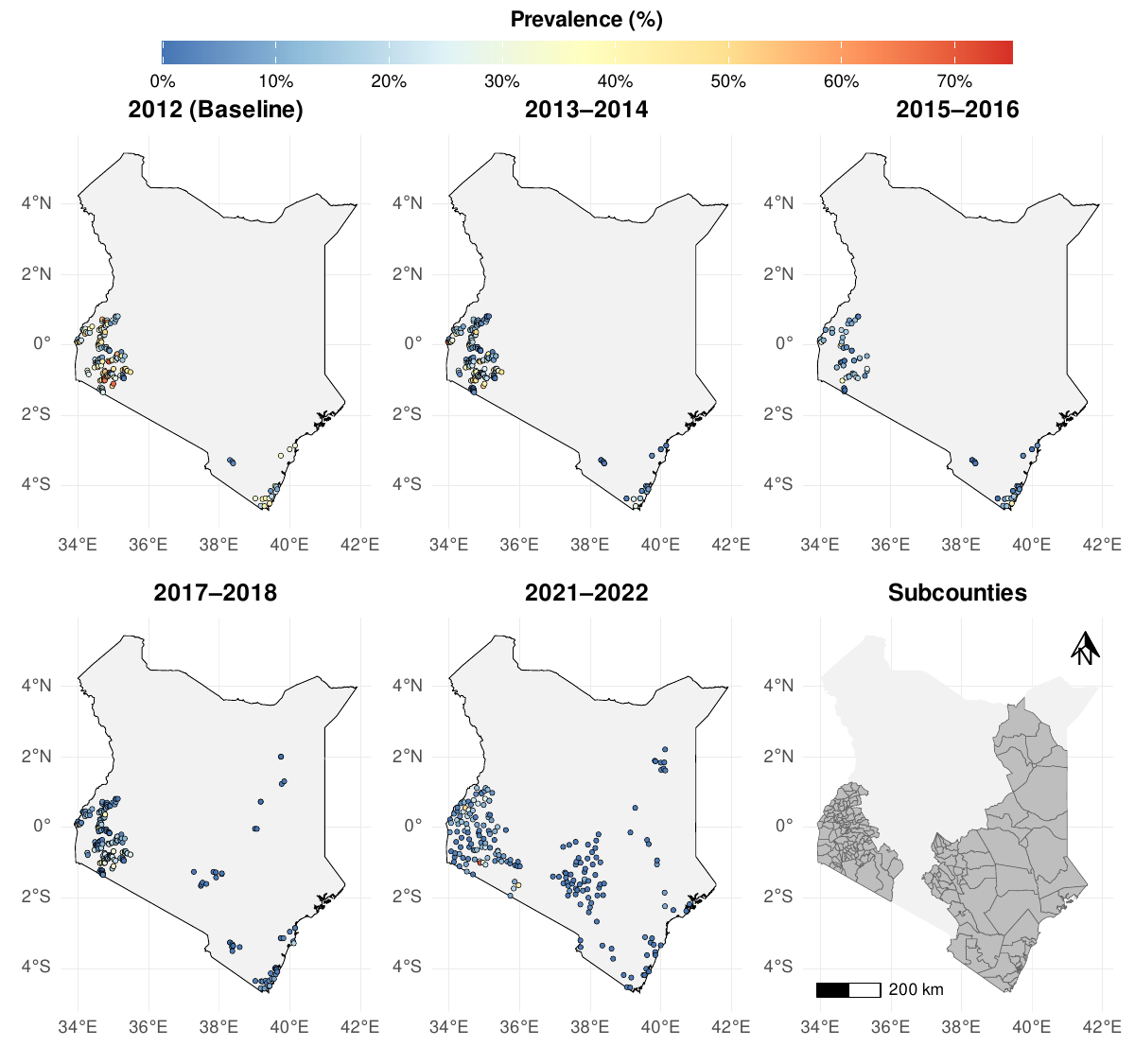}
    \caption{Spatio-temporal distribution of soil-transmitted helminth (STH) survey locations in Kenya, grouped by time period. Each panel shows the geographic distribution of surveys conducted during the specified years, with points coloured by the observed prevalence of infection with at least one STH species. The final panel displays the administrative boundaries of subcounties for which mass drug administration history data are available.}
    \label{fig:sth-data}
\end{figure}

We analyse data from STH school-based surveys conducted between 2012 and 2022 across 137 subcounties spanning 28 counties in southern Kenya; details on the sampling design used can be found in \citet{mwandawiro2019}. Species-specific infection prevalence was assessed using standardized Kato-Katz techniques. Figure~\ref{fig:sth-data} displays the survey locations alongside the boundaries of the subcounties for which we aim to predict the average prevalence for each of the three STH species.

\begin{figure}
    \centering
    \includegraphics[width=1\linewidth]{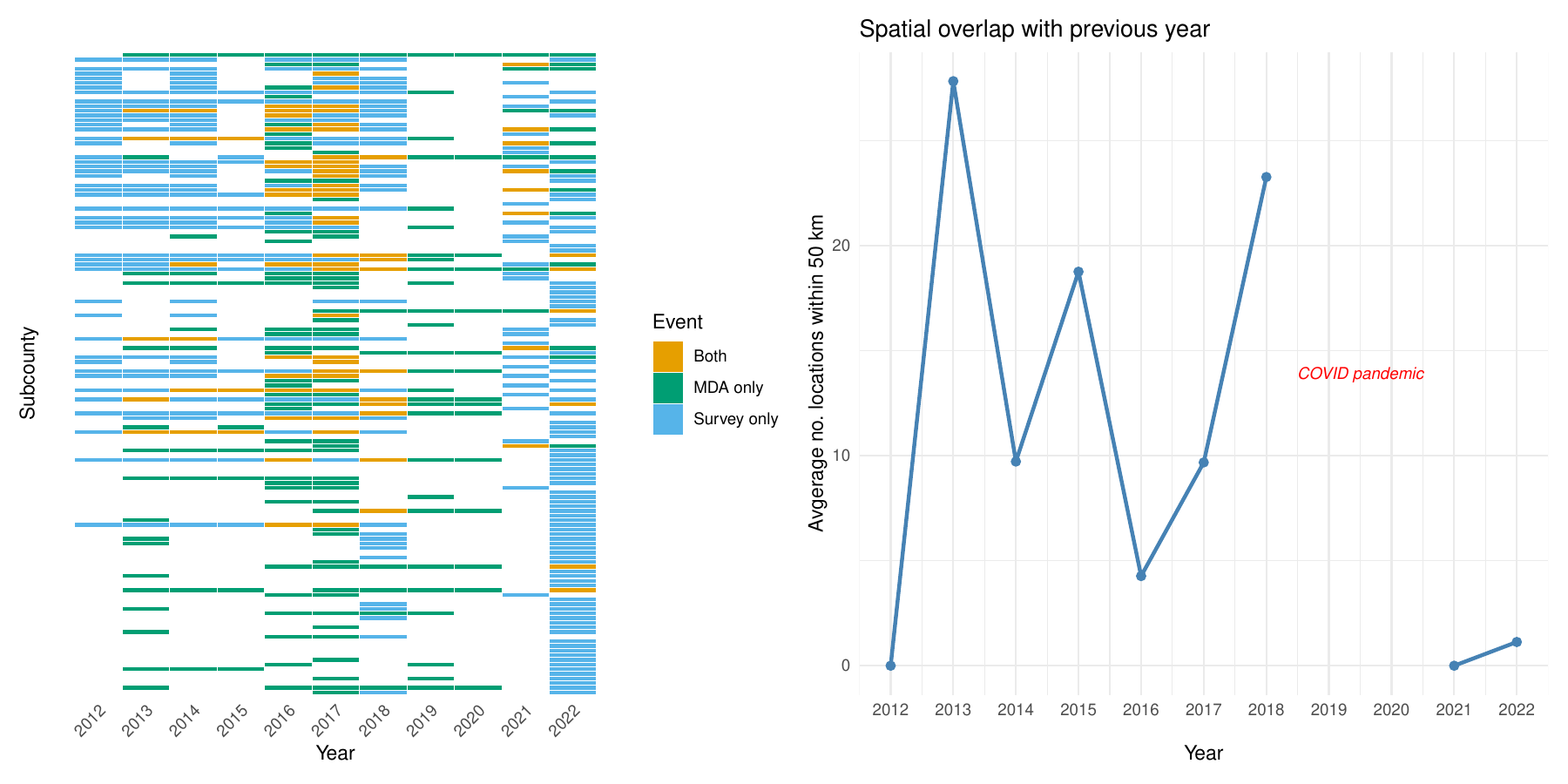}
    \caption{Left panel: Timeline of surveys and mass drug administration (MDA) activities by subcounty, with subcounties ordered from top to bottom by decreasing average prevalence of any soil-transmitted helminth (STH) infection. Orange cells indicate years where both a survey and MDA occurred, green for MDA only, and blue for survey only. Right panel: Average number of survey locations from the survey of the previous year, within a 50\,km radius. The absence of data in 2019 and 2020 corresponds to the COVID-19 pandemic, during which survey activities were disrupted.}
    \label{fig:sth-survey-mda}
\end{figure}

The left panel of Figure \ref{fig:sth-survey-mda} illustrates the timeline of survey and MDA activities across the subcounties, ordered by decreasing baseline prevalence of STH infection. Surveys, indicated by blue and orange cells, measured infection prevalence either independently or concurrently with MDA campaigns. MDA activities, shown in orange and green cells, consisted of school-based administration of albendazole, an anthelmintic drug effective against STH infections, predominantly targeting school-age children. The right panel highlights the temporal and spatial sparsity of the survey data by plotting the average number of survey locations within a 50 km radius of locations surveyed in the previous year. Data collection gaps in 2019 and 2020 reflect disruptions due to the COVID-19 pandemic. Years such as 2013 and 2018 have relatively dense spatial coverage, facilitating inference from past surveys. In contrast, more recent surveys in 2021 and 2022 exhibit limited spatial overlap, indicating a change in the areas sampled from the previous year's survey. The DAST model can therefore be especially useful here to address data sparsity and to assess the impact of MDA conducted since the baseline year of 2012.

\begin{figure}
    \centering
    \includegraphics[width=1\linewidth]{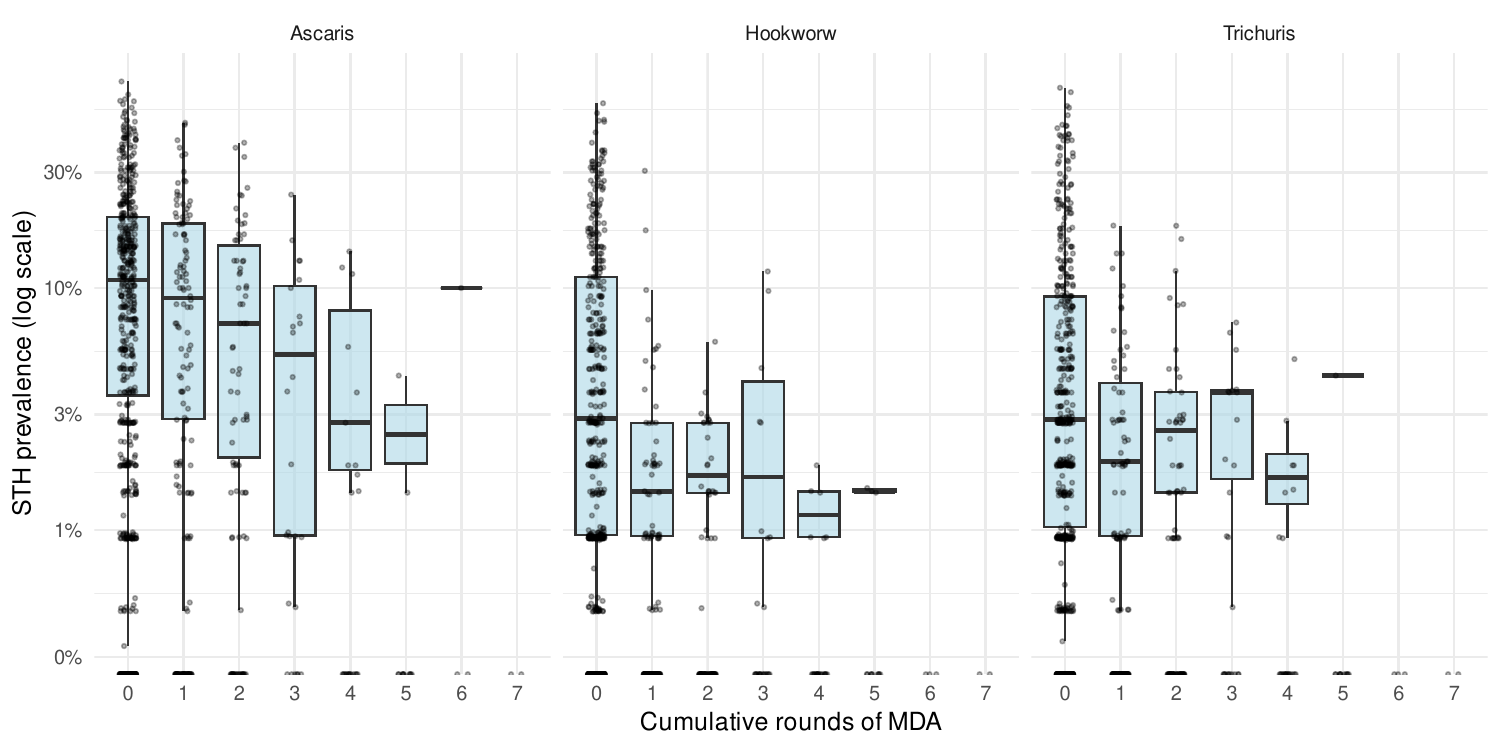}
    \caption{Boxplots of the empirical prevalence of three soil-transmitted helminths species by cumulative rounds of mass-drug-administration.}
    \label{fig:cmu_mda_sth}
\end{figure}

\begin{table}[ht]
\centering
\caption{Bias-corrected point estimates and 95\% bootstrap confidence intervals for the model parameters of each species-specific spatio-temporal model. Each entry shows the estimate followed by the lower and upper bounds of the confidence interval in square brackets. Bias correction was applied using the formula $\hat\theta_{\mathrm{BC}} = 2\hat\theta - \bar\theta_{\mathrm{boot}}$, and the intervals correspond to the 2.5th and 97.5th percentiles of the bootstrap distribution.}
\begin{tabular}{lccc}
  \hline
  Parameter & \multicolumn{3}{c}{Soil transmitted helminth species} \\
  & Ascaris & Trichuris & Hookworm \\ 
  \hline
  $\beta$ & $-2.432\;[ -2.478,\; -2.374 ]$ & $-2.402\;[ -2.457,\; -2.336 ]$ & $-1.960\;[ -2.012,\; -1.899 ]$ \\ 
  $\sigma^2$           & $22.975\;[ 11.276,\; 56.275 ]$ & $36.665\;[ 16.309,\; 90.619 ]$ & $32.760\;[ 14.418,\; 86.624 ]$ \\ 
  $\phi$               & $214.303\;[ 102.984,\; 522.853 ]$ & $221.500\;[ 99.104,\; 583.012 ]$ & $293.396\;[ 132.714,\; 844.706 ]$ \\ 
  $\alpha$             & $0.304\;[ 0.256,\; 0.371 ]$ & $0.450\;[ 0.352,\; 0.547 ]$ & $0.926\;[ 0.853,\; 0.971 ]$ \\ 
  $\gamma$             & $28.873\;[ 12.254,\; 85.485 ]$ & $3.886\;[ 2.788,\; 5.543 ]$ & $6.710\;[ 4.774,\; 8.933 ]$ \\ 
  \hline
\end{tabular}
\label{tab:bootstrap_params_sth}
\end{table}

\begin{figure}[ht]
\centering
\includegraphics[width=0.7\textwidth]{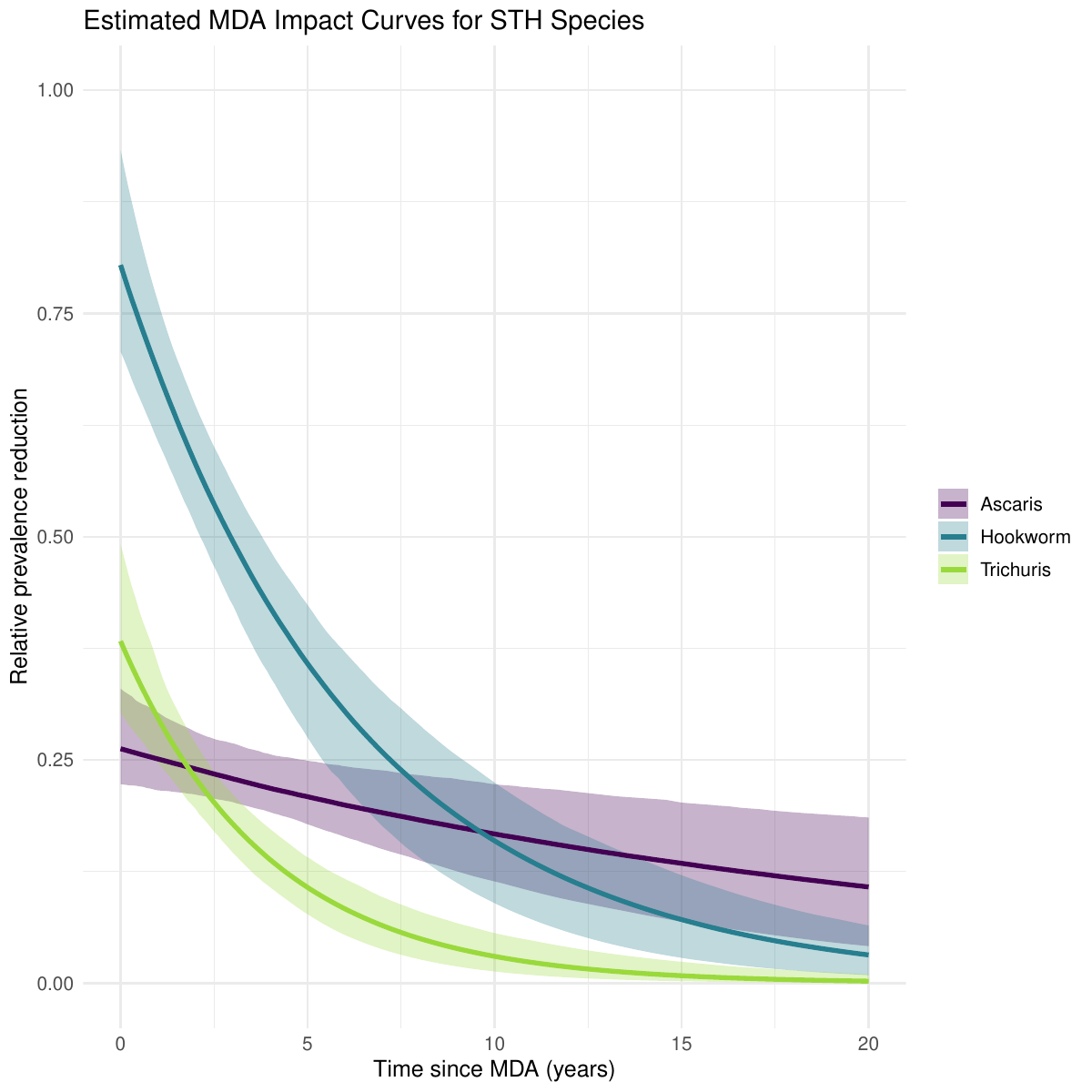}
\caption{Estimated MDA impact functions for each of the three STH species, based on the fitted exponential decay model $ f(v) = \alpha \exp(-v/\gamma) $. The shaded areas around each curve correspond to the $95\%$ confidence interval band constructed using the bootstrap samples.}
\label{fig:mda-impact-sth}
\end{figure}

Table~\ref{tab:bootstrap_params_sth} reports the point estimates and uncertainty intervals for the DAST model fitted separately to each of the three STH species, obtained using a parametric bootstrap. Following \citet{efron1979bootstrap}, new datasets are repeatedly simulated from the fitted model, the model is refitted to each simulated dataset, and the variation in the resulting parameter estimates is used to construct model-based uncertainty intervals. No penalisation for $\alpha$ was required during model fitting. The parameter estimates reveal substantial variation across species. For hookworm, the relatively high values of both $\alpha$ and $\gamma$ suggest a strong and sustained impact of MDA. In contrast, Trichuris displays a weaker and more rapidly waning response. Ascaris presents a markedly different profile, with a small immediate reduction in prevalence, $\alpha$, and a large decay parameter, $\gamma$. However, the confidence intervals for $\gamma$ in Ascaris are especially wide, reflecting substantial uncertainty in the MDA impact function. These results suggest that MDA may not have been the primary driver of reductions in Ascaris prevalence in this setting and raise concerns about the validity of the model’s assumptions for this species. However, one possible explanation is the high resilience and long environmental persistence of Ascaris eggs, which can sustain transmission even in the absence of reinfection pressure immediately following treatment \citep{senecal2020}.

Figure~\ref{fig:mda-impact-sth} shows the estimated post-MDA trajectories generated by the respective DAST models. In the absence of further treatment, it is projected that the prevalence will return to its pre-intervention levels after approximately 15 to 20 years for hookworm and trichuris. Similar multi-decade rebounds to pre-MDA transmission levels have been reported by a range of studies \citep{gunawardena2011, hardwick2019}. The species-specific patterns we observe, with more durable impact for hookworm than for ascaris, are also consistent with clinical efficacy data showing that single-dose albendazole achieves substantially higher cure and egg-reduction rates against hookworm than against trichuris, though efficacy against ascaris has been reported to be higher still \citep{moser2017}.

\subsubsection{Comparison with other geostatistical models}
\label{subsec:sth_compare}

We now compare the predictive performance of the fitted DAST model against two alternative geostatistical models, both using cumulative rounds of MDA as a covariate.

The first alternative is a binomial geostatistical model with linear predictor
\begin{equation}
\label{eq:sth_std_geo}
S: \log\left\{\frac{P(x, t)}{1 - P(x, t)}\right\} = \beta_0 + \beta_1 c(x,t) + S(x),
\end{equation}
where $c(x,t)$ represents the cumulative number of MDA rounds as defined in \eqref{eq:cum_mda}, and $S(x)$ is an isotropic and stationary Gaussian spatial process with variance $\sigma^2$ and an exponentially decaying correlation function with spatial scale parameter $\phi$. Because NTDs, including soil-transmitted helminths, are typically chronic conditions, model S, as in the standard version of DAST, assumes that all temporal variation in prevalence can be attributed to the accumulation of MDA rounds encoded in $c(x,t)$. In this scenario, prevalence exhibits a clear inverse association with $c(x,t)$, whereby larger values of $c(x,t)$ correspond to lower prevalence levels. Consequently, the use of standard geostatistical models is less problematic in this setting (Figure~\ref{fig:cmu_mda_sth}) than those described in Section \ref{sec:issues}.

The second alternative extends the first model by introducing spatio-temporal dependence through a separable, double-exponential correlation structure. This model has a linear predictor
\begin{equation}
\label{eq:sth_st_geo}
ST: \log\left\{\frac{P(x, t)}{1 - P(x, t)}\right\} = \beta_0 + \beta_1 c(x,t) + S(x, t),
\end{equation}
where $S(x,t)$ is a Gaussian process with variance $\sigma^2$, and correlation function 
$$
\text{Cor}(S(x,t), S(x',t')) = \exp\left(-\frac{\|x - x'\|}{\phi}\right) \exp\left(-\frac{|t - t'|}{\psi}\right).
$$
The inclusion of temporal correlation enables a more flexible modeling of the spatio-temporal variability than the S model by allowing for the possibility that unmeasured, time-varying factors also contribute to temporal variation in disease prevalence. 

In Table~\ref{tab:geostat_estiSth} of Appendix~\ref{app:estiSth}, we report the parameter estimates for both the S and ST models. Across all STH species, the estimated values of $\beta_1$, representing the effect of cumulative MDA rounds, are in the expected direction under both models. The plausibility and stability of these estimates suggest that
in this case the data are sufficiently informative, with no apparent issues of spatial or temporal sparsity affecting model performance.

For completeness, we also considered an extended version of the DAST model in \eqref{eq-prev} whereby the spatial random effect $S(x)$ is replaced by a fully spatio-temporal process $S^*(x,t)$.  In other words, we write
\begin{equation}
P(x, t) = P^*(x,t) \prod_{j \: : \: u_j < t} \left[1 - f(t - u_j)\right]^{\mathbb{I}(x, u_j)},
\label{eq-prev-st}
\end{equation}
and define
\begin{equation}
\log\left\{\frac{P^*(x_i, t_i)}{1 - P^*(x_i, t_i)}\right\} = d(x_i, t_i)^\top \beta + S^*(x_i, t_i),
\label{eq:endemic_st}
\end{equation}
where we have assumed a double exponential correlation function for $S^*(x, t)$ as defined in \eqref{eq:st_cor}. Here, the inclusion of $S^*(x,t)$ can be viewed as a relaxation of Assumption~A1 in Section~\ref{model}, which attributes all temporal variation to the effect of MDA and imposes a decay structure to account for temporal variation in prevalence. However, in this application, the model proved to be unidentifiable, as the MDA impact function became effectively flat, with all spatio-temporal variation absorbed by the latent process $S^*(x,t)$. This outcome reflects the empirical limitations of the available data in disentangling intervention effects from extra-Binomial spatio-temporal variation unrelated to MDA, and motivates our focus on the three models considered thus far: S, ST, and DAST.

To compare the predictive performance of alternative models, we randomly selected 10\% of the data within each survey year to construct a hold-out set. We assessed each model's calibration using the average non-randomised Probability Integral Transform (AnPIT) adapted to the geostatistical analysis of count data \citep{giorgi2021}. Specifically, for each test observation $i$, let $F_M$ denote the cumulative distribution function generated by the fitted model $M$. The non-randomised PIT is defined as
\begin{equation}
\text{nPIT}(u,y_i) = 
\begin{cases} 
0, & u \leq F_{M}(y_i-1) \\ 
\frac{u - F_{M}(y_i-1)}{F_{M}(y_i) - F_{M}(y_i-1)}, & F_{M}(y_i-1) \leq u \leq F_{M}(y) \\ 
1, & u \geq F_{M}(y_i).
\end{cases}.\end{equation}
We then take the average nPIT over the observed outcomes $y_{1}, \ldots, y_{n}$, hence $$
\text{AnPIT}(u) = \frac{1}{n} \sum_{i=1}^n \text{nPIT}(u, y_{i}). 
$$ Assessment of calibration is then carried out by checking whether the AnPIT is as close as possible to the identity function, $\text{AnPIT}(u) = u$. By plotting $\text{AnPIT}(u)$ against $u$, we can consider deviations from the identity line as evidence that the model is not well calibrated.

To complement this, we also computed the continuous ranked probability score (CRPS) for each test point $i$, defined as
\begin{equation}
\label{eq:CRPS_discrete}
\text{CRPS}(F_M,y_i)=\sum_{k=0}^{m_i}
\bigl\{F_M(k)-\mathbb{I}(k\ge y_i)\bigr\}^2,
\end{equation}
where $m_i$ is the Binomial denominator. Hence, similarly to AnPIT, we then average this over the hold out sample as
$
\sum_{i=1}^n \text{CRPS}(F_M,y_i)/n.
$
The cumulative distribution function $F_{M}$ is not available in closed form; we approximate it using Monte Carlo methods. 

\begin{figure}
    \centering
    \includegraphics[width=0.7\linewidth]{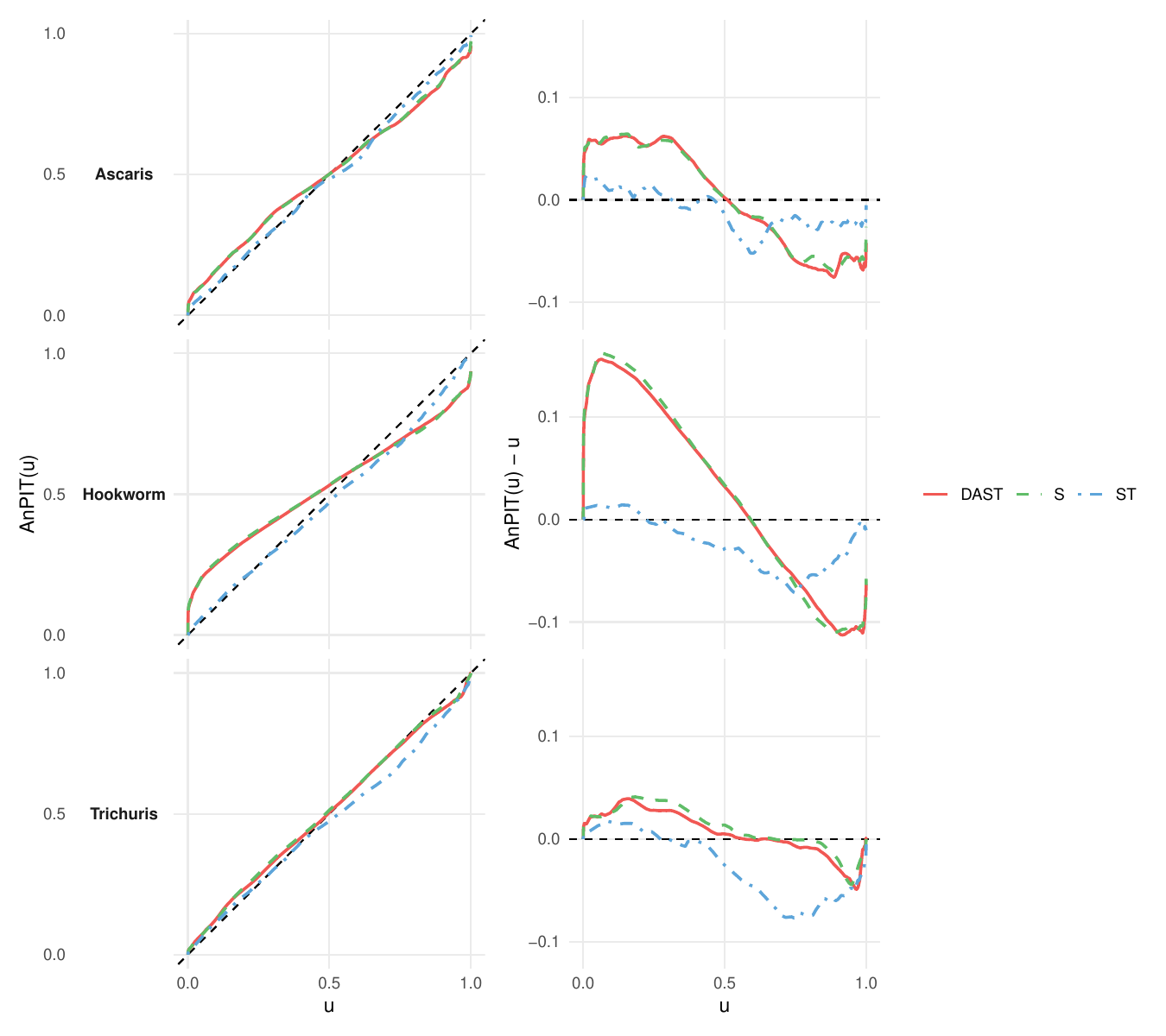}
    \caption{Average non-randomised PIT (AnPIT) curves for the held-out samples for the soil-transmitted helminths geostatistical analysis. Each row corresponds to one parasite, ascaris, trichuris and hookworm. The left column shows the AnPIT curves comparing the calibration of three models, spatio-temporal (ST, blue), spatial-only (S, green), and decay-adjusted spatio-temporal (DAST, red), with the black line representing the identity line. The right column shows the difference between each model's AnPIT curve and the identity line.}\label{fig:anpit_sth}
\end{figure}

\begin{table}
\caption{Average CRPS on the held-out data for the three models considered in the soil-transmitted helminths analysis: spatio-temporal (ST), spatial-only 
(S), and decay-adjusted spatio-temporal (DAST).
\label{tab:crps_sth}}
\centering
\begin{tabular}[t]{lrrr}
\toprule
Model  & Ascaris & Trichuris & Hookworm\\
\midrule
ST & 3.605 & 1.749 & 1.821\\
S & 3.162 & 1.214 & 3.016\\
DAST & 3.553 & 1.428 & 2.941\\
\bottomrule
\end{tabular}
\end{table}

The AnPIT plots in Figure~\ref{fig:anpit_sth} reveal some differences in calibration across the three STH species. For ascaris and hookworm, the DAST and S models exhibit similar levels of predictive calibration, while the ST model shows better alignment with the identity reference line, suggesting better calibration. In contrast, for trichuris, all three models perform comparably, with AnPIT curves indicating equally well-calibrated predictive models. As shown in Table~\ref{tab:crps_sth}, the mean CRPS values for all models and parasites are broadly similar, reflecting comparable overall predictive performance. The DAST model achieves nearly indistinguishable scores from the ST and S models, with the exception for hookworm, where the ST model provides a marginal improvement in terms of CRPS. Because both AnPIT and CRPS primarily assess probabilistic calibration, they may favour models with flexible latent processes (such as ST) even when such models obscure the effect of MDA. These scores should therefore be interpreted as complementary to, rather than substitutes for, model considerations grounded in epidemiological plausibility.

Whilst S and ST may, in some cases, offer more accurate interpolation of disease prevalence, the DAST model provides greater interpretability by explicitly linking prevalence trends to context specific motivated assumptions about the impact of MDA. As a result, DAST is better suited for programmatic applications, such as quantifying intervention effects and the impact of repeated treatment rounds. However, if the primary objective were limited to spatio-temporal interpolation, in this case study the ST model appears to be the more appropriate for ascaris and hookworm, whereas for trichuris the ST and DAST models deliver equally reliable predictions. We shall return to this point in the discussion.

\subsection{Case study II: lymphatic filariasis in Madagascar}
\label{sec:lf_mdg}

Lymphatic filariasis (LF) is a parasitic infection caused by filarial worms transmitted to humans through repeated bites from infected mosquitoes. The disease is typically acquired during childhood, but its chronic manifestations, such as lymphoedema, elephantiasis and hydrocele, usually appear later in life, resulting from long-term damage to the lymphatic system. LF is one of the leading causes of permanent disability worldwide and contributes significantly to stigma and economic hardship in affected communities. Because transmission requires sustained exposure to infective mosquito bites, control efforts have focused on large-scale, repeated rounds of MDA, which aim to reduce microfilariae in the human population and thereby interrupt the cycle of transmission. The standard treatment involves annual administration of a combination of antifilarial drugs, with regimens varying by region depending on co-endemic infections.

The present analysis focuses on lymphatic filariasis prevalence data collected in Madagascar between 2004 and 2020, comprising 579 sampled locations. These data were obtained from the Expanded Special Project for Elimination of Neglected Tropical Diseases (ESPEN) data portal developed by the World Health Organization Regional Office for Africa \citep{espen2023}. ESPEN is an open-access platform that provides epidemiological and programmatic data to support the control and elimination of neglected tropical diseases across the continent. The portal compiles and standardizes data submitted by national NTD control programmes, including results from mapping surveys, impact assessments and surveillance activities. These surveys are typically conducted at the IU level and involve community-based or school-based sampling to estimate disease prevalence and monitor the effectiveness of interventions such as mass drug administration. The ESPEN platform provides georeferenced records of prevalence estimates, treatment coverage, and intervention history, which have been widely used to map the spatial distribution of NTDs and to guide programmatic decision-making (e.g.~\cite{khaki2025}).

\begin{figure}
    \centering
    \includegraphics[width=1\linewidth]{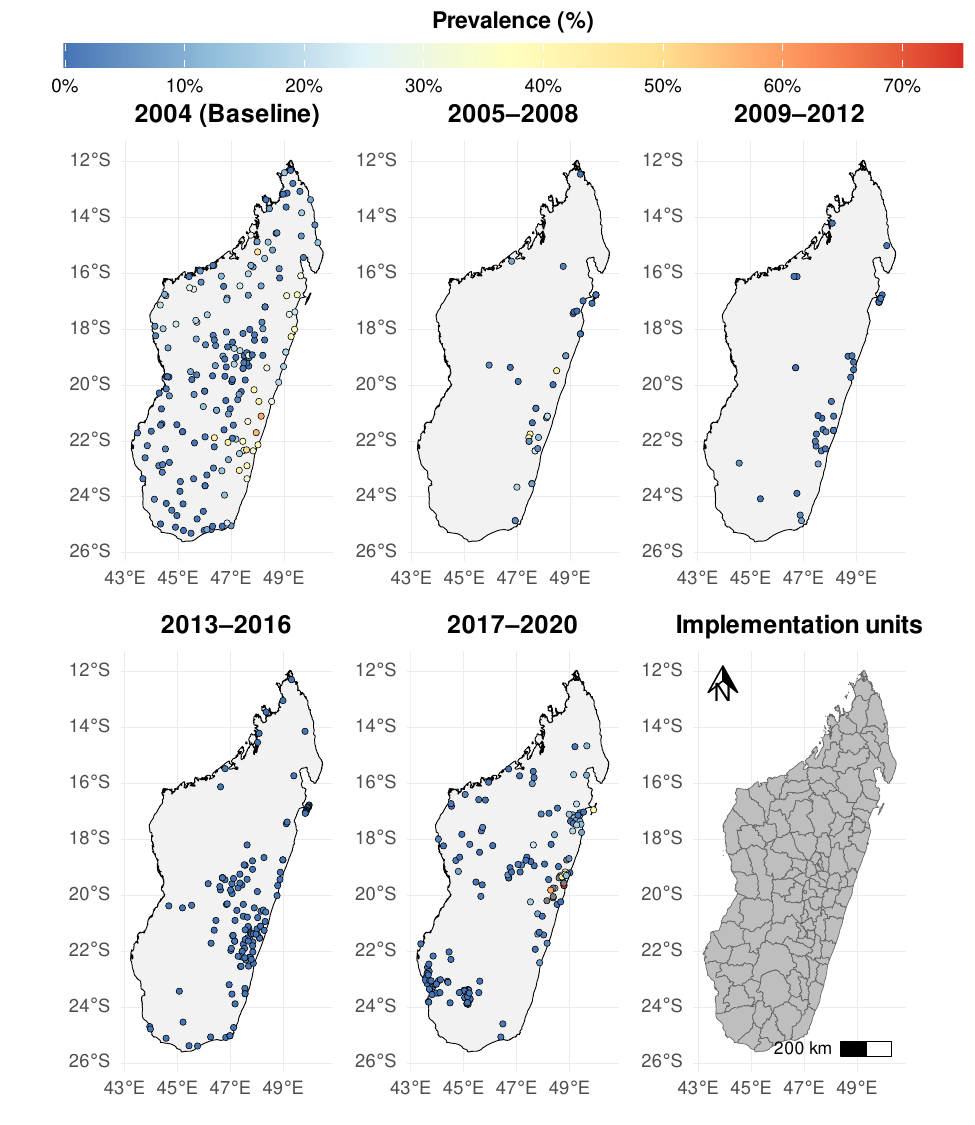}
    \caption{Spatio-temporal distribution of lymphatic filariasis (LF) survey locations in Madagascar, grouped by time period. Each panel shows the geographic distribution of surveys conducted during the specified years, with points coloured by the observed prevalence of infection with LF. The final panel displays the administrative boundaries of implementation units for which mass drug administration history data are available.}
    \label{fig:lf-data}
\end{figure}

\begin{figure}
    \centering
    \includegraphics[width=1\linewidth]{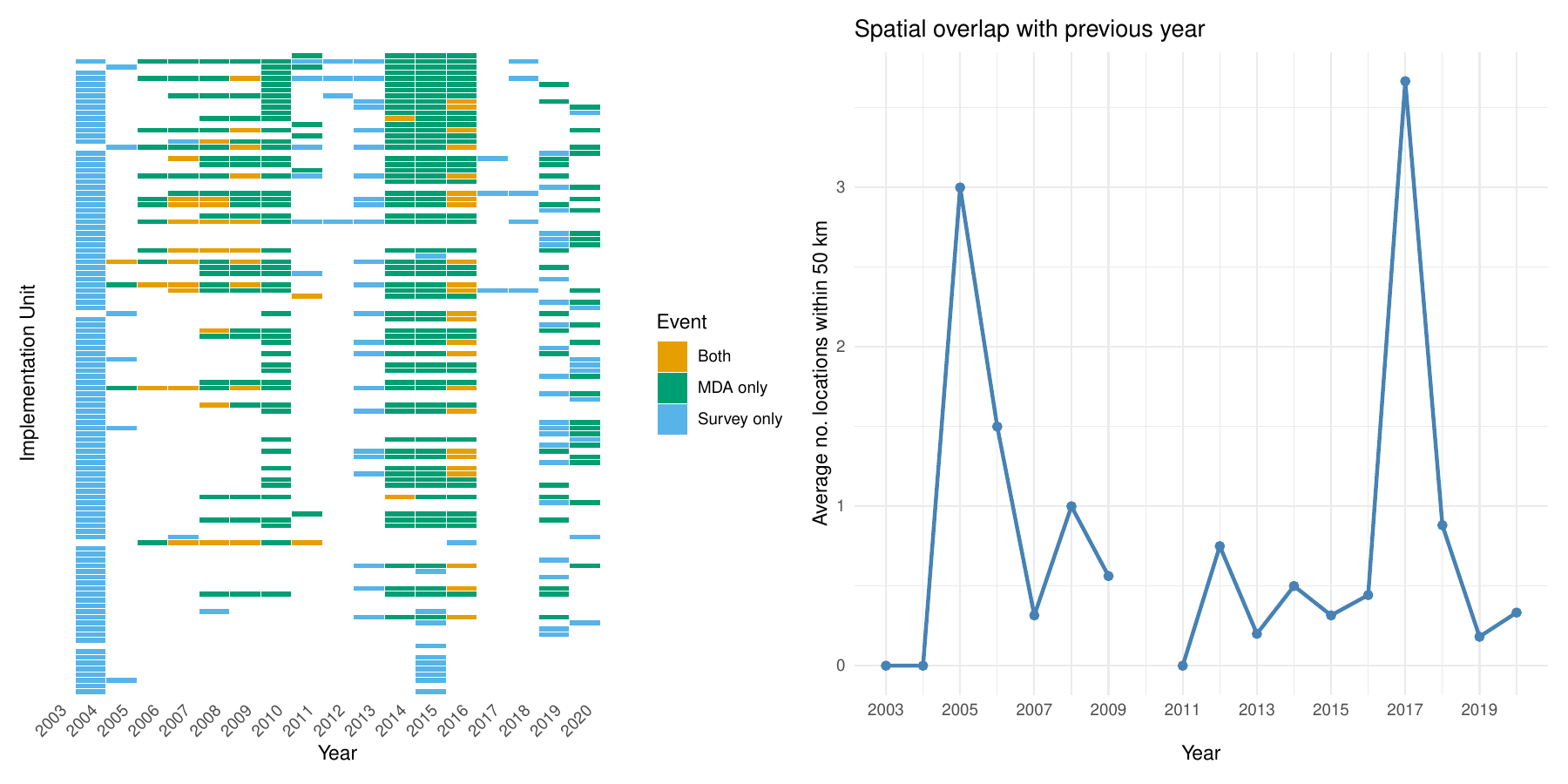}
    \caption{Left panel: Timeline of surveys and mass drug administration (MDA) activities by implementation unit (IU), with IUs ordered from top to bottom by decreasing average prevalence of lymphatic filariasis infection. Orange cells indicate years where both a survey and MDA occurred, green for MDA only, and blue for survey only. Right panel: Average number of survey locations from the survey of the previous year, within a 50\,km radius. }
    \label{fig:lf-survey-mda}
\end{figure}

LF is still endemic in Madagascar, with baseline antigenaemia prevalences exceeding $30\%$ in some southeastern districts prior to large-scale control \citep{garchitorena2018lf}. Transmission is sustained primarily by \textit{Anopheles} and \textit{Culex} mosquitoes, and although infection among children has declined, adult prevalence in certain coastal areas has been reported to exceed $10\%$ \citep{garchitorena2018lf}. Annual MDA with albendazole and diethylcarbamazine began in 2004 and was progressively expanded; by 2016, many implementation units had completed nine or more rounds \citep{garchitorena2018lf}. In 2023, the programme reached full geographical coverage for the first time, delivering treatment to over 10 million people across all 68 endemic districts via an integrated campaign \citep{who2024lf}. As shown in Figure~\ref{fig:lf-data} and Figure~\ref{fig:lf-survey-mda}, the available LF survey data for Madagascar are markedly sparse in both space and time. On average, for any given year, less than four survey locations fall within a 50\,km radius of locations sampled in the preceding year, which is considerably lower than the corresponding values for the soil-transmitted helminths dataset analysed in the previous section (see Figure \ref{fig:sth-survey-mda}). Figure~\ref{fig:lf-survey-mda} also shows that observed prevalence tends to be relatively low overall, with higher values occurring more recently, particularly during and after the COVID-19 pandemic period. This temporal pattern may be indicative of disruptions to MDA activities, which were indeed suspended or delayed in many settings due to the pandemic, potentially leading to a resurgence in infections.

\begin{figure}
    \centering
    \includegraphics[width=0.80\linewidth]{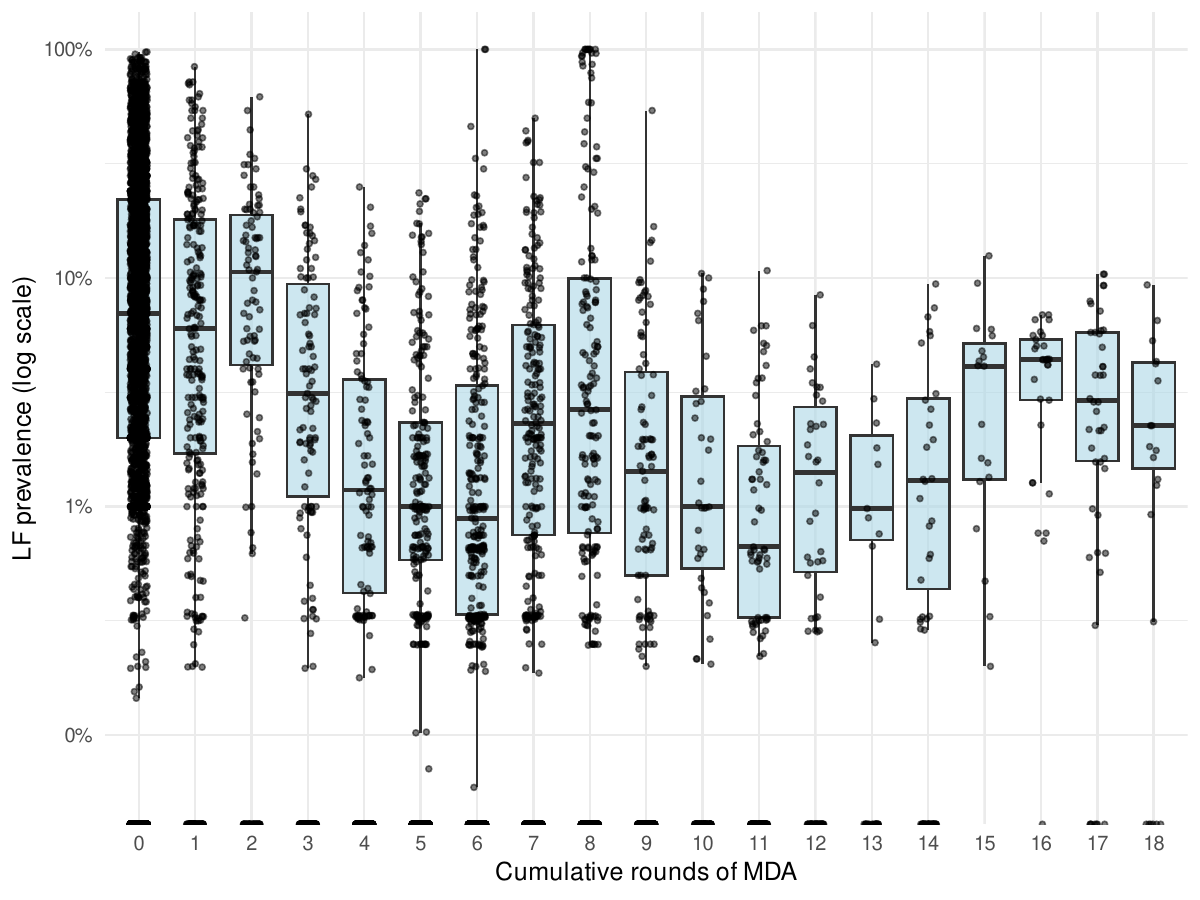}
    \caption{Boxplots of the empirical prevalence of lymphatic filariasis by cumulative rounds of mass-drug-administration.}
    \label{fig:cmu_mda_mdg}
\end{figure}

Figure~\ref{fig:cmu_mda_mdg} displays boxplots of the observed LF prevalence stratified by the cumulative number of MDA rounds received. Although there is a general trend of decreasing median prevalence with increasing rounds of MDA, particularly up to six rounds, the highest observed prevalences are nonetheless found in implementation units that have already received between seven and eight rounds. This pattern suggests that using the cumulative number of MDA rounds as a covariate may not adequately capture the true impact of treatment on prevalence, particularly since MDA has been preferentially targeted to areas with higher baseline transmission. This confounding, combined with the sparse and irregular spatio-temporal sampling noted above, complicates efforts to infer the MDA effect.

We consider three modelling approaches of increasing complexity. The first is a standard generalised linear model (GLM), in which LF prevalence $P(x_i, t_i)$ is modelled as
\begin{equation}
\label{eq:glm}
    \log\left\{\frac{P(x_i, t_i)}{1 - P(x_i, t_i)}\right\} = \beta_0 + \beta_1 c(x_i, t_i),
\end{equation}
where $c(x_i, t_i)$ denotes the cumulative number of MDA rounds (see equation \eqref{eq:cum_mda}). The second approach extends this to a generalised linear mixed model (GLMM) that includes an unstructured random effect to account for overdispersion:
\begin{equation}
\label{eq:glmm}
    \log\left\{\frac{P(x_i, t_i)}{1 - P(x_i, t_i)}\right\} = \beta_0 + \beta_1 c(x_i, t_i) + Z(x_i, t_i),
\end{equation}
where $Z(x_i, t_i)$ are independent, zero-mean Gaussian variables with variance $\sigma^2$. Finally, we implement a DAST model, in which the baseline prevalence $P^*(x_i)$ (i.e., in the absence of MDA) is modelled using an intercept-only logit-linear specification, $\log\{P^*(x_i)/(1 - P^*(x_i))\} = \beta_0$. In both the GLMM and the DAST model, we do not include a spatially correlated random effect. This choice is motivated by our empirical finding that the estimated spatial correlation is negligible, with a practical range
(the distance beyond which spatial correlation falls below 0.05) of approximately 3\,km. Given the spatial sparsity of the dataset, only around 220 pairs of locations lie within this range, which limits the ability to reliably estimate a spatial correlation structure. The
absence of broad-scale spatial correlation is not unexpected for neglected tropical diseases such as LF, which are known to exhibit highly focal transmission patterns. In the discussion, we comment on the implications of this finding for prediction of the IU-level prevalence. 

In estimating the parameters of the DAST model, the likelihood is evaluated via numerical integration. To prevent the MDA impact parameter $\alpha$ from concentrating near its upper boundary of 1, we introduce a penalisation term with $\lambda_1 = 0.35$ and $\lambda_2 = 0.35$. These values are chosen to mimic a Beta distribution with 2.5\% and 97.5\% quantiles at 0.05 and 0.95. Confidence intervals are obtained using a parametric bootstrap, which avoids reliance on asymptotic approximations and provides a more robust quantification of uncertainty under the fitted model.

Table~\ref{tab:lf-par-est} presents parameter estimates with 95\% confidence intervals for the GLM, GLMM, and DAST models, along with CRPS values computed on a 10\% hold-out sample stratified by year. There is a substantial difference between the estimates of $\beta_1$, representing the effect of cumulative MDA rounds, for the GLM and the GLMM. In the GLM, $\beta_1$ is close to zero, indicating only a weak association, whereas in the GLMM it is substantially more negative, pointing to a stronger protective effect of MDA. This contrast highlights the importance of accounting for unstructured heterogeneity across locations. Incorporating a random effect in the GLMM helps to mitigate confounding from site-level variation and yields a more plausible estimate of MDA impact. 

The parameters of the MDA impact function, $\alpha$ and $\gamma$, govern the magnitude of the initial drop in prevalence following treatment and the rate at which prevalence rebounds over time. Their point estimates imply a return to baseline levels in approximately 2.5 years in the absence of further intervention. This is about half the rebound time suggested by transmission-dynamic models, which typically estimate a return period of around 5 years or more \citep{prada2020}. While this may suggest that the DAST model underestimates the persistence of MDA effects, it is also possible that mechanistic models overestimate it, given the data-driven nature of DAST and the absence of evidence against its compatibility with the observed data. In fact, the data do not contradict the possibility of a faster rebound, which may arise from poor adherence to treatment. However, it is also important to note the wide confidence intervals for both $\alpha$ and $\gamma$, which reflect considerable uncertainty in the estimated speed of rebound. This underscores the inherent difficulty of fitting even a relatively simple DAST model to sparse survey data, and serves as a caution against introducing additional model complexity. 

\begin{figure}
    \centering
    \includegraphics[width=0.7\linewidth]{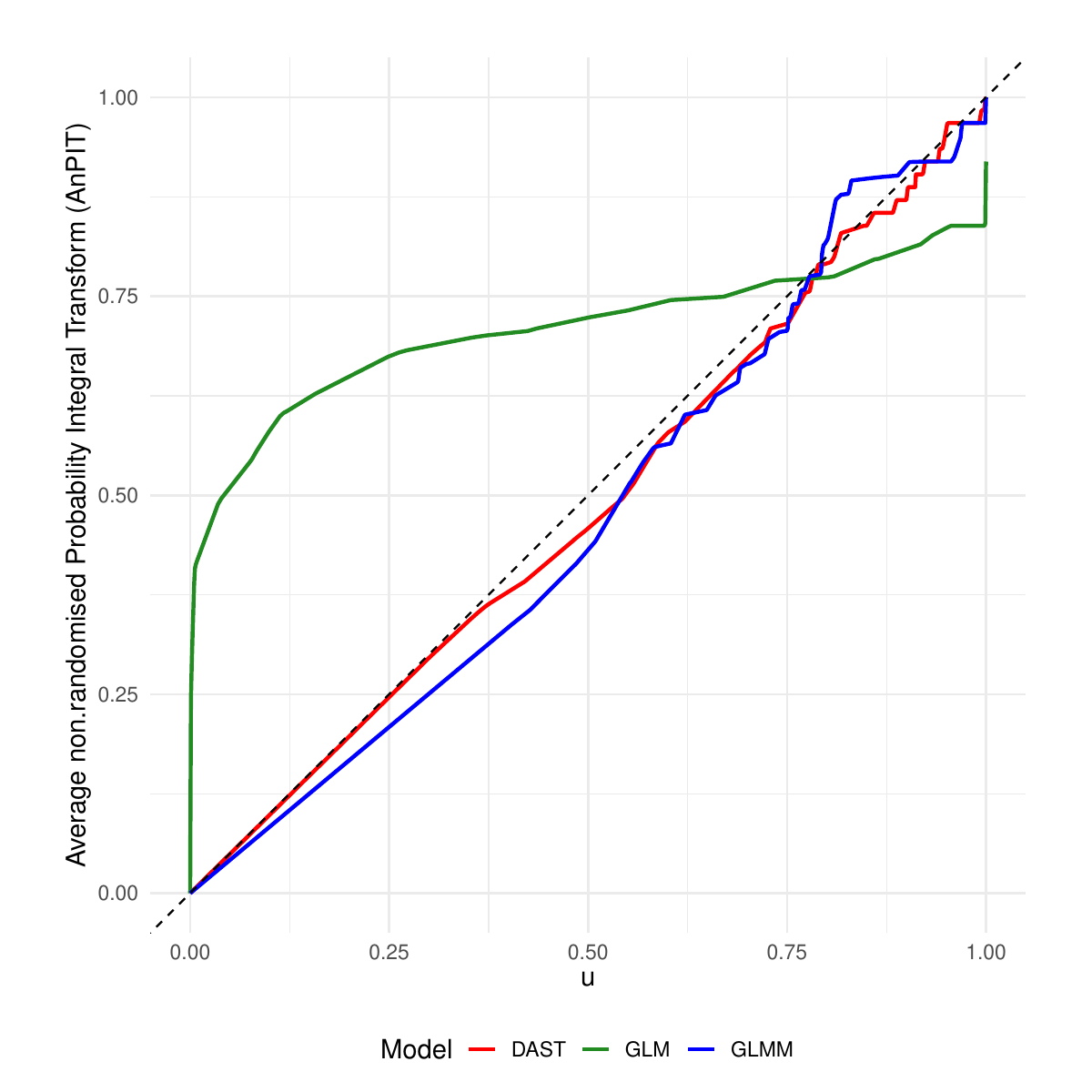}
    \caption{Average non-randomised PIT (AnPIT) curves for the held-out samples for the lymphatic filariasis data analysis. The AnPIT curves are used to assess the 
calibration of three models: Binomial generalized linear model (GLM, green), Binomial generalized linear mixed model 
(GLMM, blue), and decay-adjusted spatio-temporal (DAST, red). The black lines represent the identity line.}
    \label{fig:lf_anpit}
\end{figure}

In the DAST model, the estimated residual variance $\sigma^2$ is reduced by more than half compared with the GLMM, indicating that the adopted exponential MDA impact function accounts for a substantially greater share of the unexplained variation. By comparison with the GLMM, which relies on a simple cumulative MDA covariate, the DAST model appears to be more effective in capturing the temporal dynamics of intervention impact. Figure~\ref{fig:lf_anpit} shows the AnPIT curves for the three models. Both the GLMM and DAST curves closely follow the identity line, indicating good calibration, with the DAST model marginally outperforming the GLMM. In contrast, as expected, the GLM curve deviates substantially from the diagonal, confirming that the model is poorly calibrated due to its failure to accommodate overdispersion. To assess predictive performance, we also compute the CRPS using a hold-out sample comprising 10\% of observations randomly selected within each year. We omit the CRPS for the GLM model, as its poor calibration renders the score uninterpretable and potentially misleading. The CRPS values confirm the better performance of the DAST model, which achieves the lowest score, reflecting sharper and better calibrated predictive distributions than the GLMM.

\begin{table}[ht]
\centering
\caption{Parameter estimates with $95\%$ coverage confidence intervals, and continuous rank probability score (CRPS) for the fitted generalized linear model (GLM), generalized linear mixed model (GLMM) and decay adjusted spatiotemporal model (DAST) models to the lymphatic filariasis data from Madagascar. Confidence intervals are shown in square brackets. The CRPS are computed on a hold-out sample that is obtained by randomly selecting $10\%$ of the observations within each year.}
\label{tab:lf-par-est}
\begin{tabular}{lccc}
  \hline
Parameter & GLM & GLMM & DAST \\ 
  \hline
$\beta_0$ & -2.930 [-2.973, -2.888] & -4.217 [-4.726, -3.709] & -5.089 [-5.433, -4.740] \\ 
$\beta_1$ & -0.038 [-0.050, -0.026] & -0.875 [-0.952, -0.799] & -- \\ 
$\alpha$  & -- & -- & 0.754 [0.081, 0.823] \\ 
$\gamma$  & -- & -- & 0.441 [0.036, 5.776] \\ 
$\sigma^2$ & -- & 21.436 [17.470, 26.501] & 9.174 [7.731, 11.256] \\
  \hline
CRPS & -- & 6.303 & 5.252 \\ 
  \hline
\end{tabular}
\end{table}

\begin{figure}
    \centering
    \includegraphics[width=0.6\linewidth]{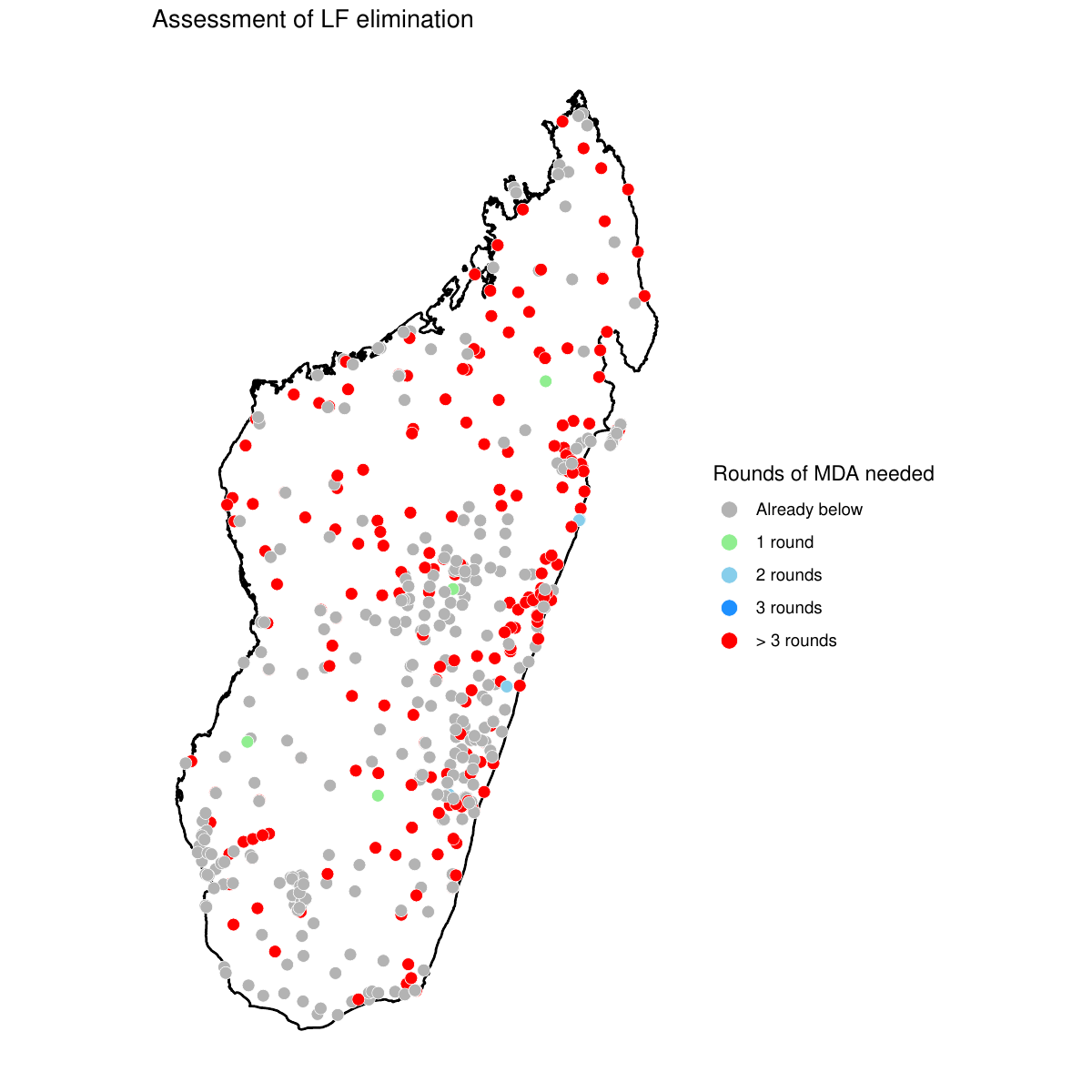}
    \caption{Predicted number of additional annual mass drug administration (MDA) rounds required to reduce lymphatic filariasis prevalence below $1\%$ at surveyed locations in Madagascar. Estimates are based on posterior predictive means from the decay adjusted spatio-temporal model, assuming up to three consecutive MDA rounds administered annually from 2021 onwards, following the most recent survey year in 2020.}
    \label{fig:lf-rounds}
\end{figure}

In light of the better performance and good calibration of the DAST model, we use it to estimate, for each surveyed location, the number of additional annual MDA rounds required to bring LF prevalence below the 1\% threshold for elimination, following the last year of data collection in 2020. Figure~\ref{fig:lf-rounds} reveals a striking contrast between locations where prevalence is already projected to be below 1\% in 2020, and those where up to three further rounds are predicted to be necessary. Only five locations are projected to require one additional round, and just four would need two, underscoring the highly focal nature of LF transmission. This focality likely reflects a complex interplay of ecological, demographic, and operational factors that shape local transmission dynamics and the effectiveness of past interventions.

\section{Discussion}
\label{sec:discussion}

We have introduced a novel geostatistical modeling framework based on a decay-adjusted spatio-temporal (DAST) model specifically designed to quantify the impact of mass drug administration (MDA) on the prevalence of neglected tropical diseases (NTDs). The DAST model approximates the temporal dynamics of disease prevalence following MDA using an exponential decay function. One of the main advantages of the proposed approach over standard methods, where MDA rounds are typically included as a simple regression covariates, is that DAST fully leverages the history of MDA by accounting for both the timing and cumulative effect of successive treatments. A key component of the DAST model is the MDA impact function, which we parameterize as $f(v) = \alpha \exp\{-v/\gamma\}$, where $v$ denotes the time since MDA, $\alpha$ represents the immediate proportional reduction in prevalence following MDA, and $\gamma$ controls the duration of this effect. By explicitly modelling the decay in prevalence following MDA, we have shown that DAST enables reliable inference even under challenging data scenarios, where spatial and temporal sparsity, combined with the targeted roll-out of MDA in high-risk areas, can otherwise lead to biased estimates of the cumulative effect of multiple MDA rounds and undermine the utility of MDA data for predicting disease prevalence. DAST provides a data-driven approach to address programmatic questions related to both the mapping of NTD prevalence and the evaluation of alternative MDA strategies for disease control. Hence, DAST offers modelling flexibility that makes it applicable across a range of NTDs and adaptable to diverse epidemiological settings, as illustrated by the two case studies presented in this paper.

Although the motivating applications in this paper are drawn from NTDs, the proposed DAST framework
is not inherently disease-specific and may
therefore be applied more broadly to settings in which outcomes respond to repeated or time–varying interventions under sparse
and temporally misaligned observation schemes. The key modelling principle underpinning DAST is the decomposition of the observed outcome into a baseline
component and a cumulative intervention effect whose
influence evolves over time. This structure naturally extends to a wide range of applications, including the evaluation of vaccination programmes, the impact of
other public‑health interventions (e.g.   insecticide‑treated bed nets), and non‑epidemiological contexts such as changes in service utilisation following policy interventions. Importantly, while the general framework is transferable, the  specification of the intervention impact function $f(\cdot)$ and the assumptions governing how
effects accumulate and decay over time must be informed by domain‑specific knowledge and data availability. In particular, the exponential decay adopted here reflects biological considerations and the coarse temporal resolution typical of routine NTD survey data; alternative functional forms may be more appropriate in contexts where intervention effects are delayed,  partially permanent, or governed by threshold dynamics. We therefore view DAST as a flexible modelling template rather than a prescriptive formulation, whose components should be adapted to the scientific and operational context of each application.

\subsection{Extending the DAST modelling framework}
\label{sec:extesions}

Our modelling framework relies on a set of core assumptions that approximate the effect of MDA on disease prevalence. In this section, we revisit each of the assumptions outlined in Section~\ref{model} and discuss potential extensions to the current framework. These extensions, which we view as promising directions for future research, aim to relax some of the simplifying assumptions while preserving the interpretability and computational tractability of the model.

First, we have assumed baseline equilibrium (A1), whereby, in the absence of intervention, prevalence at any location fluctuates around a constant mean. This assumption may not hold in the presence of substantial population mobility. For instance, movement of individuals between areas with differing transmission intensities can lead to the importation of new infections, thereby disrupting local equilibrium and introducing transient fluctuations in prevalence that are unrelated to intervention effects. Moreover, the assumption of baseline equilibrium may also be undermined by interventions targeting other diseases that share ecological or transmission pathways with the NTD of interest. A notable example is the distribution of insecticide-treated bed nets for malaria control, which can simultaneously reduce transmission of lymphatic filariasis in areas where both diseases are transmitted by \textit{Anopheles} mosquitoes. Such cross-intervention effects introduce additional sources of temporal variation in prevalence that are not attributable to MDA alone, and thus warrant careful consideration when modelling disease dynamics. A natural extension to address this limitation would be to replace the purely spatial Gaussian process $S(x)$ with a spatio-temporal process $S(x, t)$, thereby allowing for background changes in prevalence over time that are independent of MDA. However, as demonstrated in the application of Section~\ref{sec:sth_kenya}, this approach can be empirically problematic, since the added flexibility of $S(x, t)$ may absorb much of the variation that would otherwise be attributed to the impact function $f(v)$, leading to spatio-temporal confounding and weakening the identifiability of MDA effects. In such instances, particularly when prior information is available for the parameters governing $f(v)$, a possible solution is to regularize the estimation of the impact function directly. In a fully Bayesian framework, this can be achieved by placing informative priors on the parameters $\alpha$ and $\gamma$. Alternatively, in a penalized likelihood setting one may apply regularization terms to each of these parameters, similar to the approach already illustrated in this paper. This has the advantage of allowing targeted regularization of the MDA impact function without requiring prior specification for the remaining model parameters, for which informative priors are typically unavailable.

We have assumed that each round of MDA produces an immediate relative reduction in prevalence that subsequently wanes if no further rounds are administered (A2). This assumption may be more tenable for some NTDs than others. For infections with rapid reinfection dynamics and weak or short-lived immunity, such as soil-transmitted helminthiases, schistosomiasis, and trachoma, the effect of a single round of treatment commonly diminishes over relatively short time scales, making the transient-impact formulation a reasonable approximation. By contrast, for vector-borne infections such as lymphatic filariasis and, to a lesser extent, onchocerciasis, repeated MDAs can drive transmission below a disease-specific breakpoint (or push the effective reproduction number below one), after which prevalence may not rebound even in the absence of continued MDA. In these settings, the waning to baseline implied by assumption A2 may be overly pessimistic during the later phases of programmes approaching elimination. From a modelling perspective, this motivates extensions to the impact function $f(v)$ to allow for partial permanence of the MDA effect. One such extension is to define the impact function as
\begin{equation}
f(x, t) = \alpha \left[(1 - r(x,t)) \exp\left\{-\frac{t - t^*(x,t)}{\gamma}\right\} + r(x,t)\right],
\end{equation}
where $t^*(x,t)$ denotes the time of the most recent MDA prior to time $t$ at location $x$. The function $r(x,t) \in [0,1]$ represents the fraction of the MDA-induced reduction that persists indefinitely. To reflect the cumulative effect of repeated interventions, $r(x,t)$ can be modelled as a monotonically increasing function of the cumulative number of MDA rounds $c(x,t)$. For instance, a saturating function may be used to ensure that $r(x,t) \to 1$ as $c(x,t)$ increases, while remaining close to zero in areas with limited intervention history. This formulation captures the idea that sustained MDA pressure progressively reduces the likelihood of full rebound. Alternatively, threshold-based rules can be applied to specify the onset of partial permanence. Specifically, we define $f(v, x,t) = \alpha \Big[(1 - r(x,t)) \exp\{-v/\gamma\} + \delta r(x,t)\Big]$,
$$
r(x,t) = \mathbb{I}\{P(x, t^-) < p^\ast\},
$$
where $P(x, t^-)$ is the latent prevalence at location $x$ immediately before time $t$, $p^\ast \in (0,1)$ is a pre-defined elimination threshold, and $\delta \in (0,1)$ is an unknown parameter controlling the fraction of the initial reduction that is retained once the threshold is crossed. In this formulation, prevalence initially decays according to the transient component of the impact function, but once it drops below $p^\ast$, a proportion $\delta$ of the reduction is preserved indefinitely. This allows the model to represent a shift from transient to partially permanent MDA impact.

We have assumed that MDA induces a constant relative reduction in prevalence, such that the treatment effect acts multiplicatively and does not depend on the pre-intervention level (A3). While this simplifies the modelling framework, it may not always hold in practice due to several biological and programmatic factors. For instance, systematic non-adherence by individuals who never participate in MDA, heterogeneous coverage patterns that correlate with baseline prevalence, and density-dependent parasite biology can all result in a relative reduction that varies with the underlying prevalence. In some cases, drug efficacy may be intensity-dependent. For example, treatment may be more effective against high parasite loads because a larger number of parasites increases the likelihood that a sufficient proportion are exposed to the drug at therapeutic concentrations. In addition, individuals with heavier infections may exhibit stronger immune activation following drug-induced parasite death, resulting in enhanced clearance. Conversely, in low-intensity infections, partial exposure or lower immune engagement may lead to reduced treatment efficacy. To account for these effects, a natural extension is to allow the immediate proportional reduction $\alpha$ to depend on the pre-intervention prevalence $P^*(x)$. Hence, we could replace the constant $\alpha$ with
$$
\alpha(x) = h\!\left(P^*(x);\, \boldsymbol{\theta}\right),
$$
where $h(\cdot)$ is a smooth, bounded function mapping \([0,1] \to (0,1)\), and $\boldsymbol{\theta}$ are parameters to be estimated. The function $h$ can be interpreted as a data-driven dose-response relationship that links the pre-MDA prevalence to the expected relative reduction achieved by treatment. The shape of $h$ would thus reflect how efficacy varies across different epidemiological contexts. For example, $h$ may be increasing if treatment is more effective in high-prevalence areas, either because drug action is more complete when parasite loads are high, or because such settings typically receive higher coverage or adherence due to greater perceived disease burden. Conversely, $h$ may be decreasing if efficacy diminishes at higher prevalence levels, which could arise from saturation effects (e.g., reduced per-parasite drug exposure), lower per-person adherence in heavily affected areas, or immune system fatigue. Suitable choices for $h$ include monotonic continuous functions or, alternatively, $\alpha$ may also be modelled as a piecewise-constant function across prevalence strata:
$$
\alpha(x) = \alpha_j \quad \text{if } P^*(x) \in I_j,
$$
where $I_j$ is a partition of the prevalence domain \([0,1]\) into discrete intervals (e.g., low, moderate, high), and each \(\alpha_j\) is a separate parameter estimated for class $j$. This categorical specification can be especially useful when data are sparse or when prior programmatic thresholds for intervention intensity exist. Such extensions would allow the DAST model to reflect realistic variations in treatment effect across different epidemiological contexts.

Finally, we have assumed that impact surveys are conducted after the MDA effect has reached its peak, so that each observation lies on the decay part of $f(v)$ (A4). This greatly aids identifiability of the decay parameters but is often untestable with routine programme data for which survey dates are coarsely spaced, for example by month or quarter, MDA rounds may be delivered over windows of days or weeks, and the time-to-peak itself is unknown. Relaxing A4 without more granular temporal information
post-MDA risks non-identifiability between the peak height, the rise-to-peak dynamics, and the decay rate, because the same observed prevalence can be explained by “early, still-rising” or “late, already-decaying” post-MDA trajectories. Our view is that, in the absence of repeated measurements shortly after treatment, it is not possible to disentangle the rise-to-peak phase from the subsequent decay and A4 can only be relaxed if strong prior information is available on the timing of the peak effect of MDA. Nevertheless, we regard this assumption as less critical than the others discussed above. The aim of the DAST model is not to estimate the timing or height of the peak effect, which would require longitudinal data, but to characterise the rate at which the MDA effect wanes over time.

Other extensions, beyond those discussed above can also be considered, depending on the research question and the availability of supporting data. For example, if the aim is to quantify programme effectiveness in populations with variable adherence, to model heterogeneity across age groups, or to evaluate the combined effect of multiple overlapping interventions, the basic structure of the model can be adapted accordingly. Below we describe three such extensions.

Firstly, if adherence data are available as estimated proportions of individuals in each location who actually swallowed the drug during each MDA round, these can be incorporated directly into the multiplicative structure of the model. In the original formulation, the impact of each round is assumed to be a fixed proportion $\alpha$, and prevalence declines as a function of the number of rounds $c(x,t)$. However, in reality, the effectiveness of each round depends not only on delivery but also on uptake. Let $A(x, t)\in[0,1]$ be adherence at location $x$ and time $t$; hence, we write
$$
P(x,t) = P^*(x)\prod_{j:\,u_j < t} \Big(1 - f(t-u_j)\Big)^{A(x, u_j) I(x, u_j)} .
$$
When $A(x, t)=1$ for all $x$ and $t$, this reduces to the original formulation. Values of $A(x,t)$ less than one reduce the effect of poorly adhered rounds and allow the model to reflect intervention intensity more accurately.

Secondly, treatment effects may vary systematically with age. Age-specific heterogeneity in treatment response can arise from multiple factors: younger individuals often have higher parasite burdens, which may lead to greater absolute reductions following treatment; pharmacokinetics and pharmacodynamics may differ between children and adults; and adherence patterns are typically age-structured, with school-based delivery targeting specific age groups. These differences can be captured by 
allowing both the immediate reduction and the rate of decay to depend on age, $a$,
replacing the standard impact function $f(v)$ with
$$
f(v, a) = \alpha(a) \exp\left\{-\frac{v}{\gamma(a)}\right\}.
$$
Here, $\alpha(a)$ and $\gamma(a)$ could be modelled as smooth functions (e.g. using splines) or as piecewise constants over age strata.

Third, in settings where multiple interventions are being implemented, such as MDA and vector control, the multiplicative model can be extended to account for each intervention’s contribution separately. A convenient way to accommodate this is to retain the multiplicative structure and assign each intervention its own impact function and decay profile:
$$
P(x,t) = P^*(x)\prod_{k=1}^K \prod_{j:\,u_{k,j}<t} \big\{1 - f_k(t-u_{k,j})\big\}^{\mathbb{I}_k(x,u_{k,j})},
$$
where $k$ indexes the intervention type, $u_{k,j}$ is the time of its $j$th administration, and $f_k$ may take the same exponentially decaying form as in DAST but with intervention-specific parameters $(\alpha_k,\gamma_k)$. When two interventions $k$ and $\ell$ are co-delivered, potential \emph{synergy} or \emph{antagonism} can be captured through an interaction term. A simple formulation is:
$$
f_{k\ell}(v) = 1 - \big(1 - f_k(v)\big)\big(1 - f_\ell(v)\big)\big(1 - \psi_{k\ell}\big),
$$
where $\psi_{k\ell} > 0$ indicates synergy (i.e., the combined effect is greater than multiplicative), $\psi_{k\ell} < 0$ indicates antagonism (i.e., overlapping or interfering effects), and $\psi_{k\ell} = 0$ recovers the standard multiplicative model. However, reliable estimation of the interaction parameter $\psi_{k\ell}$ is likely to be challenging without strong prior information, and a data-structure that includes temporal staggering, spatial heterogeneity in intervention deployment, or both.

\subsection{The balance between model complexity and empirical identifiability}
\label{sec:model_complexity}

All of the extensions discussed above are appealing in principle, as they allow the model to reflect biological mechanisms and programmatic realities  more closely. However, each additional layer of complexity increases the difficulty of estimation, particularly in the sparse and heterogeneous data settings that are common in NTD programmes. As we have shown, even the basic DAST model can be challenging to estimate in such contexts. For this reason, in this paper we have chosen to focus on the simplest formulation, leaving a more detailed investigation of the proposed extensions to future work.

More generally, adding spatio-temporal random effects $S(x,t)$, age-varying parameters $\alpha(a)$ and $\gamma(a)$, adherence-weighted MDA impacts, or multiple-intervention interaction terms can all be justified by specific scientific or programmatic objectives. However, these should only be introduced once three conditions are met: (1) the simpler model is poorly calibrated; (2) the additional component is identifiable, for instance through simulation-based calibration, prior sensitivity analysis, or cross-validated predictive checks; and (3) the parameters of interest can be estimated with sufficient precision to support reliable decision-making. In the absence of these conditions, one risks fitting models whose estimates are driven more by assumptions than by the information present in the data. This emphasis on empirical identifiability distinguishes our approach from traditional transmission-dynamic or mathematical modelling frameworks, which are often used in NTD programme planning and forecasting; see, for
example \cite{truscott2016soil, prada2020}. Such models rely on mechanistic assumptions about infection dynamics and intervention effects, and are typically fitted using a combination of partial data and strong priors derived from the literature or expert opinion. While this enables long-term forecasting in the absence of rich data, it also makes inferences heavily dependent on the assumed model structure and parameter values. By contrast, the DAST framework seeks to extract information from the available data with minimal reliance on external assumptions, favouring a statistical approach that prioritises identifiability and uncertainty quantification.

Our case study on lymphatic filariasis in Madagascar (Section~\ref{sec:lf_mdg}) illustrates the importance of parsimony. Here, we found negligible spatial correlation and therefore used a non-spatial DAST mixed model. If the goal had been to estimate prevalence at the implementation-unit level, two alternative approaches could have been considered. One is to extend the non-spatial DAST model by including spatially referenced covariates to capture broader variation. This approach relies on covariates that are truly predictive of prevalence, which may be difficult in low-prevalence settings. Another option is to adopt a design-based approach, where the predictive distribution from the DAST model is used to estimate prevalence at the sampled locations. This can then be combined with design-weighted direct estimators, model-assisted estimators, or small-area estimation methods, as discussed by \citet{wakefield2020}. However, this too presents challenges: some IUs may include too few observations to support reliable estimation. In short, both strategies may necessitate additional data and do not overcome the fundamental limitations posed by the sparse sampling designs typical of NTD surveys.

\section{Conclusions}
\label{sec:conclusions}

We have proposed a decay-adjusted spatio-temporal (DAST) model that explicitly accounts for the time-varying impact of mass drug administration on the prevalence of neglected tropical diseases (NTDs), providing an interpretable, data-driven framework suitable for spatially and temporally sparse survey settings. Through applications to soil-transmitted helminths and lymphatic filariasis, we have demonstrated that DAST offers a feasible alternative to standard geostatistical models, particularly when the goal is to quantify the impact of interventions or to produce short-term forecasts for programmatic decision making. The framework is broadly applicable to a range of NTDs and other environmentally-driven infections. Future work will explore model extensions that better capture biological and programmatic complexities, and establish identifiability conditions under which such models can be reliably estimated.

\begin{appendices}

\section{Parameter estimates for the standard geostatistical models of Section \ref{sec:sth_kenya}}
\label{app:estiSth}

\begin{table}[htbp]
\centering
\caption{Maximum likelihood estimates with 95\% Wald confidence intervals, for the spatial (S, from equation \eqref{eq:sth_std_geo}) and spatio–temporal (ST, from equation \eqref{eq:sth_st_geo}) geostatistical models for the application on soil-transmitted helminths of Section \ref{sec:sth_kenya}.}
\label{tab:geostat_estiSth}
\resizebox{\textwidth}{!}{%
\begin{tabular}{llcccccc}
\toprule
 & & \multicolumn{2}{c}{\textit{Ascaris}} & \multicolumn{2}{c}{\textit{Trichuris}} & \multicolumn{2}{c}{Hookworm} \\
\cmidrule(lr){3-4} \cmidrule(lr){5-6} \cmidrule(lr){7-8}
Parameter &  & S & ST & S & ST & S & ST \\
\midrule
$\beta_0$ &
& $-3.900$ ($-4.494$, $-3.305$)
& $-5.672$ ($-6.325$, $-5.019$)
& $-4.978$ ($-5.065$, $-4.890$)
& $-5.988$ ($-8.033$, $-3.944$)
& $-5.605$ ($-6.390$, $-4.819$)
& $-6.046$ ($-6.905$, $-5.187$) \\
$\beta_1$ &
& $-0.384$ ($-0.418$, $-0.350$)
& $-0.129$ ($-0.275$, $0.017$)
& $-0.258$ ($-0.261$, $-0.256$)
& $-0.082$ ($-0.104$, $-0.060$)
& $-1.137$ ($-1.222$, $-1.051$)
& $-0.119$ ($-0.164$, $-0.074$) \\
$\sigma^2$ &
& 6.820 (5.259, 8.836)
& 4.957 (3.634, 6.774)
& 4.113 (3.979, 4.251)
& 5.279 (2.877, 9.685)
& 5.303 (4.002, 7.026)
& 5.336 (3.379, 8.423) \\
$\phi$ &
& 59.307 (41.173, 85.451)
& 41.391 (29.401, 58.266)
& 22.014 (20.723, 23.379)
& 27.531 (14.812, 51.181)
& 41.888 (26.553, 66.040)
& 59.243 (33.931, 103.440) \\
$\psi$ &
& --- & 11.056 (8.598, 14.197)
& --- & 20.881 (13.297, 32.793)
& --- & 6.615 (3.956, 11.041) \\
\bottomrule
\end{tabular}%
}
\end{table}

\end{appendices}

\section{Competing interests}
No competing interest is declared.

\section{Acknowledgments}
 This work was carried out under the Geostatistics for Global Health project, funded by the Gates Foundation (grant no. INV-095716). We would like to thank Evidence Action's Deworm the World Initiative and the Kenya Medical Research Institute (KEMRI) for obtaining and providing the Kenya data set. We are also grateful to the Associate Editor and the two referees for their constructive comments, which have led to a substantially improved manuscript.

\section{Data availability}
The data analysed in this study are not publicly available. Requests for access may be made to the corresponding author. All code and R scripts used for the simulation study and case studies are publicly available on GitHub at \url{github.com/giorgistat/dast-paper}.

\section*{Abbreviations}

\begin{list}{}%
  {%
    \setlength{\leftmargin}{2.8cm}%
    \setlength{\labelwidth}{2.4cm}%
    \setlength{\labelsep}{0.4cm}%
  }
  \item[AnPIT] Average non-randomised Probability Integral Transform
  \item[CRPS] Continuous Ranked Probability Score
  \item[DAST] Decay-Adjusted Spatio-Temporal
  \item[EVI] Enhanced Vegetation Index
  \item[ERR] Egg Reduction Rate
  \item[ESPEN] Expanded Special Project for Elimination of Neglected Tropical Diseases
  \item[GLM] Generalised Linear Model
  \item[GLMM] Generalised Linear Mixed Model
  \item[IU] Implementation Unit
  \item[LF] Lymphatic Filariasis
  \item[MALA] Metropolis-Adjusted Langevin Algorithm
  \item[MBG] Model-Based Geostatistics
  \item[MCMC] Markov Chain Monte Carlo
  \item[MCML] Monte Carlo Maximum Likelihood
  \item[MDA] Mass Drug Administration
  \item[NDVI] Normalised Difference Vegetation Index
  \item[NTD] Neglected Tropical Disease
  \item[PC] Preventive Chemotherapy
  \item[PIT] Probability Integral Transform
  \item[RMSE] Root Mean Square Error
  \item[SE] Standard Error
  \item[ST] Spatio-Temporal
  \item[STH] Soil-Transmitted Helminths
  \item[WHO] World Health Organization
\end{list}

\bibliographystyle{abbrvnat}
\bibliography{biblio}



\end{document}